\documentclass[letterpaper, 10 pt, conference]{ieeeconf} 
\usepackage{caption}
\usepackage{subcaption}
\usepackage{graphicx}
\usepackage{siunitx}
\usepackage{textcomp} 
\usepackage{amsmath,amssymb}
\usepackage{float}
\usepackage{physics}
\usepackage{comment}
\IEEEoverridecommandlockouts 

                                \newcommand*\diff{\mathop{}\!\mathrm{d}}     

\title{\LARGE \bf
Neural Fiber Activation in Unipolar vs Bipolar Deep Brain Stimulation*
}

\author{Anna Franziska Frigge $^{1}$, Alexander Medvedev $^{1}$, Elena Jiltsova $^{2}$, and Dag Nyholm $^{2}$
\thanks{*This work is part of the project ``Patient-specific dynamical modeling and optimization of deep brain stimulation" funded within The EU Joint Programme – Neurodegenerative Disease Research 
by the Swedish Research Council, Grant 2020-02901.}
\thanks{$^{1}$ Information Technology,
        Uppsala University, SE-75236 Uppsala, Sweden
        {\tt\small \{anna.frigge,alexander.medvedev\}@it.uu.se}}
\thanks{$^{2}$ Neurology, Uppsala University Hospital, SE-75185 Uppsala, Sweden
        {\tt\small \{elena.jiltsova,dag.nyholm \}@neuro.uu.se}}
}

\begin{document}
\maketitle
\thispagestyle{empty}
\pagestyle{empty}

\begin{abstract}
Deep Brain Stimulation (DBS) is an established and powerful treatment method in various neurological disorders. It involves chronically delivering electrical pulses to a certain stimulation target in the brain in order to alleviate the symptoms of a disease.
Traditionally, the effect of DBS on neural tissue has been modeled based on the geometrical intersection of the static Volume of Tissue Activated (VTA) and the stimulation target. Recent studies suggest that the Dentato-Rubro-Thalamic Tract (DRTT) may serve as a potential common underlying stimulation target for tremor control in Essential Tremor (ET). However, clinical observations highlight that the therapeutic effect of DBS, especially in  ET, is strongly influenced by the dynamic DBS parameters such as pulse width and frequency, as well as stimulation polarity. This study introduces a computational model to elucidate the effect of the stimulation signal shape on the DRTT under neural input.  
The simulation results suggest that achieving a specific pulse amplitude threshold is necessary before eliciting the therapeutic effect through adjustments in pulse widths and frequencies becomes feasible. Longer pulse widths proved more likely to induce firing, thus requiring a lower stimulation amplitude. Additionally, the modulation effect of bipolar configurations on neural traffic was found to vary significantly depending on the chosen stimulation polarity and the direction of neural traffic. Further, bipolar configurations demonstrated the ability to selectively influence firing patterns in different fiber tracts.
\end{abstract}

\section{INTRODUCTION}
The technology of Deep Brain Stimulation (DBS) has critically advanced our capacity to influence neural circuits and provide significant  relief of symptoms in neurological disorders, e.g. Essential Tremor (ET). In DBS, an electrode delivering chronic stimulation is surgically implanted in the targeted brain  area that is selected depending on the symptoms.
Yet, the therapeutical effect of DBS is contingent on the  stimulation settings that are at present manually chosen by medical personnel based on symptom response and individualized through a trial-and-error tuning procedure.
This paper suggests  computational tools for the quantification of neural fiber activation in DBS, thus paving the way to {\it in silico} individualization of the DBS settings via mathematical optimization. 

DBS can be viewed as a control system, with stimulation parameters serving as inputs and symptom severity as the controlled output~\cite{Medvedev2019}. Motor neurological  symptoms, e.g. tremor, are conventionally assessed through clinical evaluations conducted by medical personnel, but can also be measured using e.g. smartphones or smartwatches~\cite{Olsson2020}. The precise neural mechanisms underlying the therapeutic effect of  DBS remain unclear. There are three primary theories currently under debate: (i) the inhibition of neural signals, (ii) the excitation of neural signals, and (iii) the disruption of abnormal information flow~\cite{Chiken2016}. Notably, the model framework introduced in this paper can be employed to explore all the considered scenarios.

\paragraph*{Stimulation target} Traditionally, the ventralis intermediate nucleus (Vim) was the primary DBS target in ET. Over time, the potential of other targets, e.g. the caudal zona incerta (cZI) and the subthalamic nucleus, have been investigated \cite{Chandra2022,Holslag2018}. The superiority of one target over the others has been widely debated, but is not yet proven unequivocally. Nevertheless, several studies indicate that the heterogenity of stimulation targets in ET might be explainable by a shared underlying pathway, the Dentato-Rubro-Thalamic Tract (DRTT)~\cite{Yang2020,Dembek2020,Middlebrooks2021}. A recent study by Middlebrooks \textit{et al.}~\cite{Middlebrooks2021} confirmed a strong correlation between tremor improvement and stimulation of the DRTT. Moreover, it was found that several previously reported stimulation sweetspots overlap the DRTT \cite{Middlebrooks2021}. At the same time, quantifying the effect of the stimulation on neural fiber remains an open issue. For instance, no correlation between improved tremor and the distance to the DRTT is reported in \cite{Schlaier2015}.

The DRTT consists of a larger decussating (dDRTT) and a smaller non-decussating part (ndDRTT)~\cite{Petersen2018}. Deuter \textit{et al.}~\cite{Deuter2022} suggest that both fiber bundles are correlated to a good clinical response, but that the effect of the dDRTT is more prominent.

\paragraph*{Electrical DBS model} Modeling the spatial electric field distribution is a common approach to capture the effect of the DBS stimulation on neural tissue. Static models, which are disregarding the pulsatile character of the DBS signal, are widely utilized to calculate the Volume of Tissue Activated (VTA) due to their simplicity. 
Static VTA computations usually involve the utilization of the second spatial derivative of the extracellular potential (activating function)~\cite{Butson2005} or, more commonly, the electric field norm~\cite{Astrom2015}, which quantifies the magnitude of the electric field generated by the DBS electrodes. Specifically, the computation of the VTA often relies on identifying instances where the electric field norm surpasses a predefined threshold.
While VTA models can be useful in predicting the effect of DBS~\cite{Butson2007,Cubo2019}, they aggregate large neural populations exhibiting complex connectivity and  impulsive dynamics into tissue characterized by a certain conductivity value. The complex nature of neural firing, which is strongly influenced by various factors such as ion channel dynamics, network dynamics, and synaptic plasticity, is neglected. Thus, it is important to highlight that VTA models only provide a coarse approximation of the stimulation threshold required to elicit a neural response.

Another drawback of the present VTA models is, that they fail to capture the effect of dynamical DBS parameters, e.g. pulse width and frequency. Moreover, there is a lack of consensus on how to include bipolarity in VTA models. Duffley \textit{et al.}~\cite{Duffley2019} demonstrated that different activating functions and the electric field norm can produce similar VTAs for unipolar settings, whereas they can differ significantly for bipolar settings. 

Modeling the impact of bipolar DBS configurations is especially crucial, given that these have demonstrated comparable or superior effectiveness to unipolar settings, while potentially offering advantages such as the reduction of adverse side effects~\cite{Steffen2020} and the extension of battery life~\cite{Almeida2016}.

\paragraph*{Neuron models} Neuron cable models simulate the electrical properties of neurons and can be used to study how electrical signals propagate through them. Various neuron models~\cite{McIntyre2002,Astrom2015} that take the effect of extracellular stimulation via a DBS electrode into consideration have been developed. Moreover, there have been several attempts to correlate the results of neuron model simulation under dynamical DBS settings to electric field distributions from static VTA models. In particular, Åström \textit{et al.}~\cite{Astrom2015} investigated how cable diameter and pulse width affect the neural response produced by a cable model placed at different locations relative to the lead. The paper provides VTA threshold values for different pulse widths and cable diameters, thus enabling static VTA models to capture the effect of dynamical parameters. However, only unipolar settings were taken into consideration, the electric field distributions were based on a homogeneous FEM model without encapsulation layer, and the  orientation of the neuron relative to the lead was neglected. 

In this paper, the impact of stimulation configurations  and settings on the fibers of the dDRTT and ndDRTT are examined. To this end, a cable model is positioned near the lead at specific locations within the chosen fiber tracts. 

The  main contributions of the study are as follows. 

\begin{enumerate}
    \item \textit{Dynamic modeling scenario:}
     The concept of modifying a traversing axonal current and assessing the DBS effect in terms of phase shift~\cite{Andersson2018} has been extended to a more realistic scenario. This involved placing the lead at a position relative to the neuron model that aligns with actual clinical situations and portraying a realistic fiber geometry. This approach is patient-specific and makes use of an actual orientation of the cable model relative to the lead. 
    \item \textit{Effect of dynamic stimulation parameters:} The results highlight the necessity of reaching a specific amplitude threshold before modulation via pulse width and frequency adjustments can be achieved. Moreover, longer pulse widths were found to promote firing, whereas the effect of different frequencies was less distinct.
    \item \textit{Bipolar stimulation efficacy:} Flipping the polarity in a bipolar DBS configuration distinctly affects neural traffic. Moreover, bipolar settings demonstrated the ability to selectively impact firing patterns within the dDRTT and ndDRTT.
\end{enumerate}
To the best of our knowledge, this study is the first one that explores and quantifies the effects of DBS pulses in a realistic fiber tract scenario taking  into account a traversing axonal current. Further, it emphasises  bipolar stimulation settings, which are frequently overlooked in the literature due to the complexity and challenges associated with accurately measuring their effects in computational models.
The findings align well with clinical observations and lay the basis for a computational tool predicting the outcome of a stimulation signal adjustment in DBS.

The rest of the paper is organized as follows. First, the modeling set-up and simulation scenario are outlined, providing a detailed description of the FEM model and cable model utilized. Moreover, a firing score to capture the binary effect of firing is introduced. Then, the simulation results are presented and discussed.
\section{Computational modeling}
The simulation scenario focuses on investigating the effect of DBS on brain tissue near the lead, with an emphasis  on neural traffic within the DRTT fiber tracts. A train of four consequent DBS pulses impacting the neural dynamics are considered. For simplicity, only a single axonal input pulse is taken into account, thus assuming a significantly lower frequency of neural traffic compared to the DBS frequency. 

Typically,  stimulation from a DBS lead implanted in one hemisphere has a more pronounced effect on symptoms in the contralateral side (e.g., a lead in the left hemisphere affecting a right hand tremor). Nevertheless, clinical practice often reveals cases where a single lead influences both sides. Such bilateral effects may be attributed to the stimulation of both the dDRTT and ndDRTTs~\cite{Middlebrooks2021}. Consequently, this study investigates the effects on both the dDRTT and ndDRTT to shed light on this phenomenon.

\begin{figure}
      \centering
      \includegraphics[angle=0,origin=c,scale=0.12]{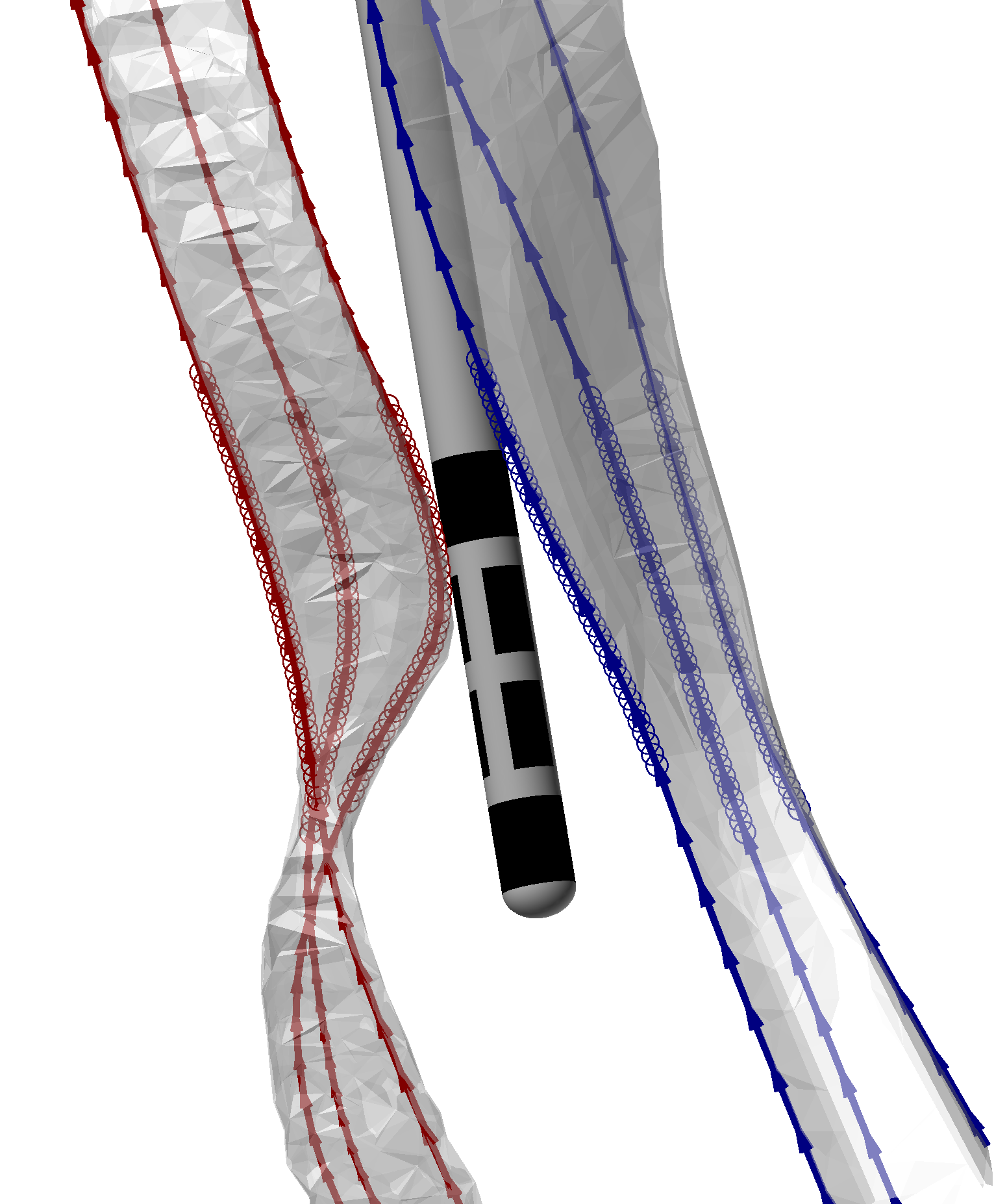}
      \caption{The positioning of the DBS lead with respect to both the dDRTT (red) and the ndDRTT (blue). Three individual fibers are highlighted in each tract. Tract coordinates were derived from an atlas~\cite{Middlebrooks2021}, whereas the lead coordinates originate from an actual patient. Circles mark the locations where the FEM solution was evaluated.}
      \label{fig:lead_and_tracts}
   \end{figure}

The modeling set-up consists of a FEM model of the DBS lead and neuron models. First, the  FEM model  with time-varying boundary conditions defined by the stimulation signal is employed to solve for the electric potential distribution generated by the DBS pulses. Second, the electric potential is fed into the neuron models positioned at the coordinates corresponding to fibers within the dDRTT and ndDRTT. Fig.~\ref{fig:lead_and_tracts} depicts the positioning of the lead relative to the two parts of the DRTT. The FEM model and the neuron model are presented in detail in the subsequent sections.

\subsection{FEM model}
\begin{figure}
      \centering
      \includegraphics[scale=0.17]{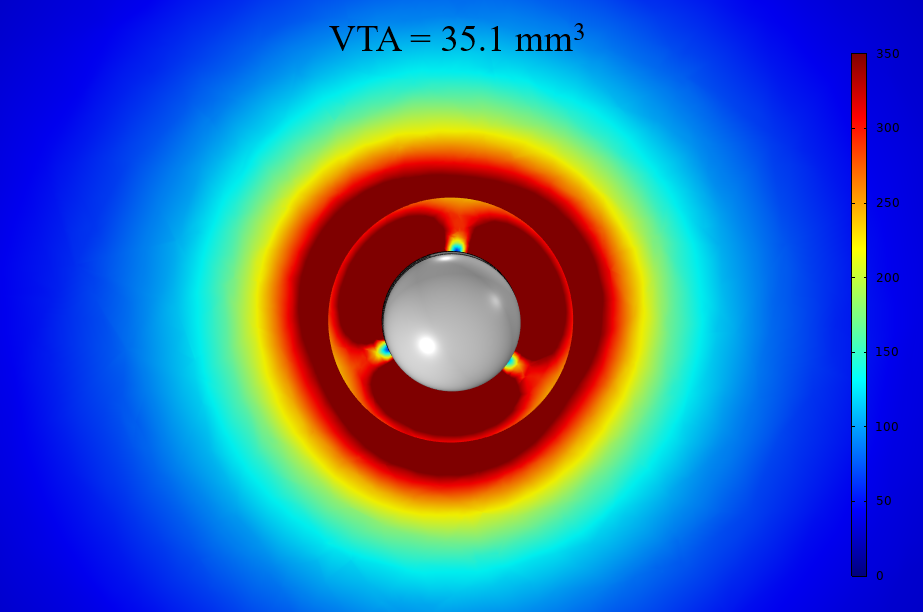}
      \includegraphics[scale=0.17]{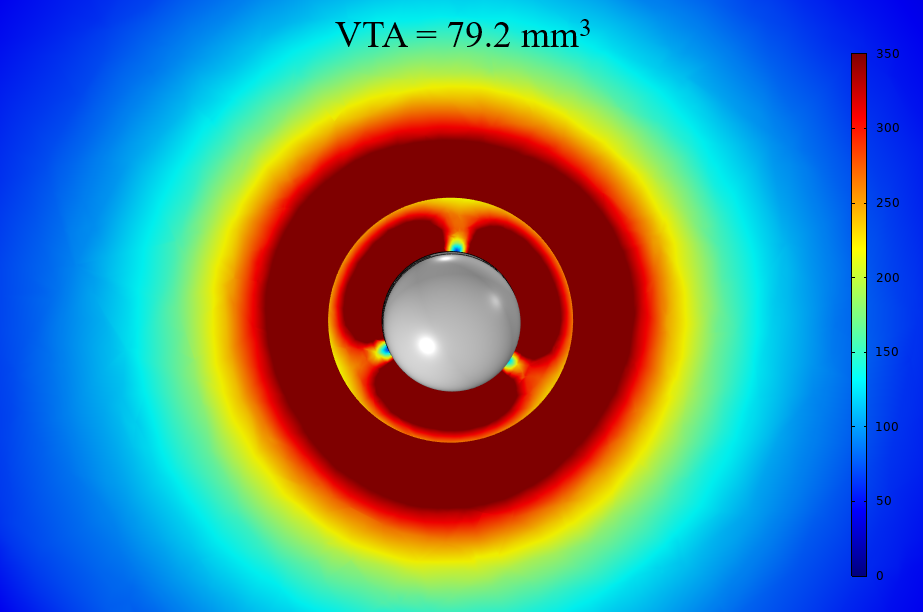}
      \caption{Electric field norm for the FEM model with homogeneous (left) and heterogeneous (right) conductivity. Notably, the transition from the homogeneous model, where conductivity is assumed to be constant, to the heterogeneous model, which incorporates varying conductivities for gray matter, white matter, and cerebrospinal fluid, results in the VTA more than doubling. The different tissue types around the lead were derived from segmented MRI atlas data warped to the patient's anatomical space.}
      \label{fig:conductivity}
   \end{figure}

\begin{figure}
  \centering
  \begin{subfigure}[b]{0.25\textwidth}
    \centering
    \includegraphics[width=\textwidth]{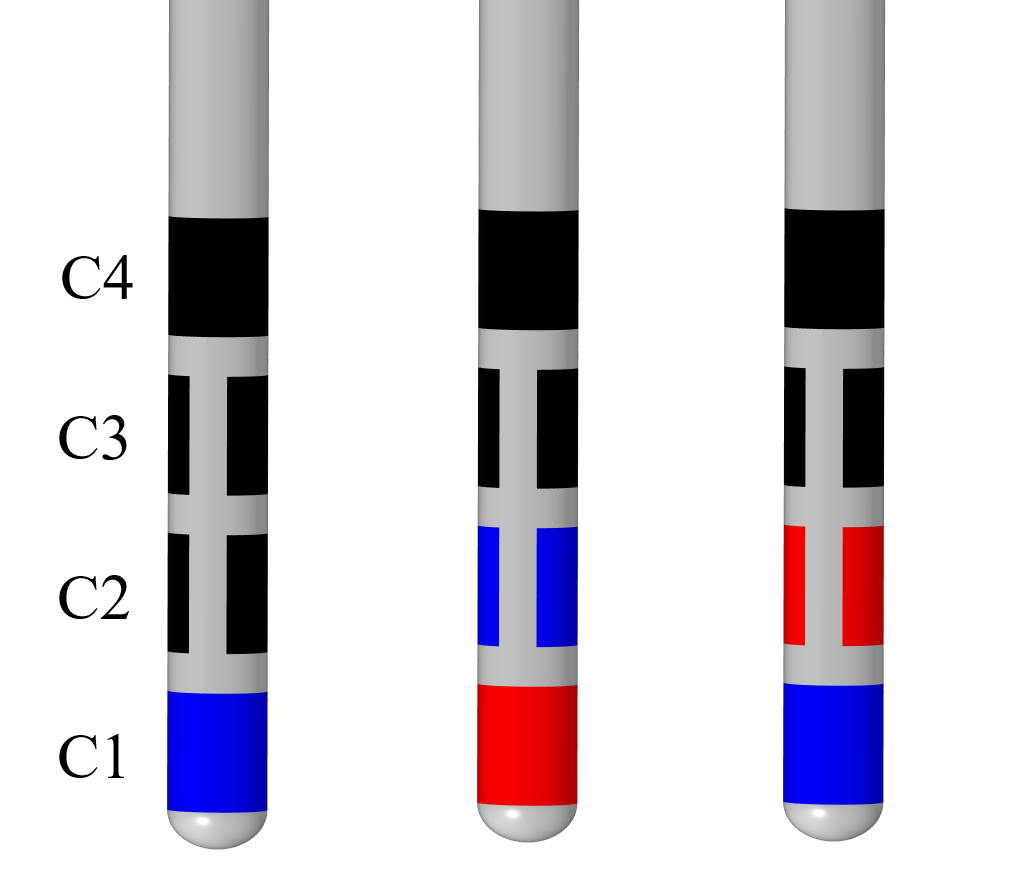}
    \caption{}
  \end{subfigure}
  \hfill
  \begin{subfigure}[b]{0.45\textwidth}
    \centering
    \includegraphics[width=\textwidth]{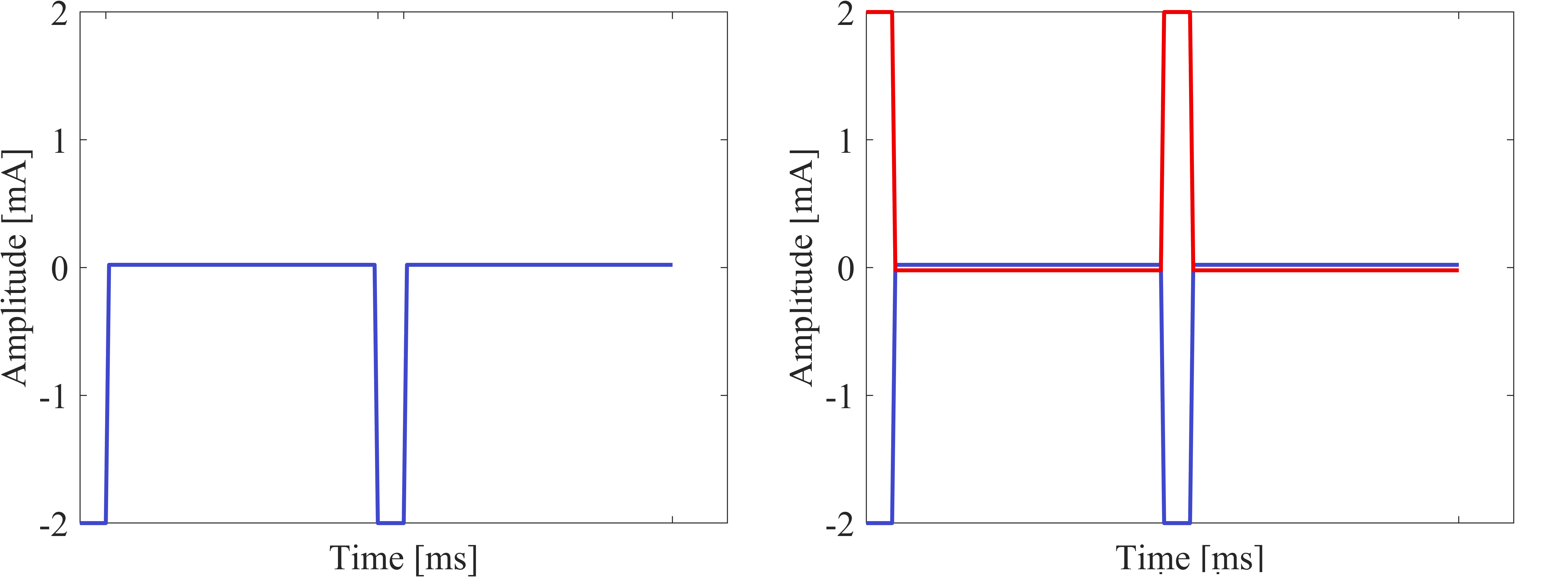}
    \caption{}
  \end{subfigure}
  \caption{(a) Active contacts in uni- and bipolar configuration settings. Cathodic stimulation is in blue, anodic stimulation is  in red. From left to right: (i) a unipolar cathodic (C1-), (ii) a bipolar (C1-, C2+), and (iii) the bipolar configuration with reversed polarity (C2-, C1+).
  In (b), two DBS pulses are plotted for a unipolar cathodic stimulus (left) and a bipolar stimulus (right).}
  \label{fig:uni_bipolar}
\end{figure}

A FEM model was created in COMSOL Multiphysics 5.6. to assess the propagation of the electric field generated by the stimulation through the brain tissue. The model solves the following linear partial differential equation in a 3-D space and across time:
\begin{equation}
    \nabla \cdot \mathbf{J} = -\nabla \cdot (\sigma \nabla u+\epsilon_0 \epsilon_r \nabla u) =0. 
    \label{eq:pde}
\end{equation}
Here, $\nabla \cdot$ denotes the divergence,  $\mathbf{J}$ represents the current density, $\sigma$  is the conductivity, and $\nabla u$ stands for the gradient of the electric potential. The constants $\epsilon_0$ and $\epsilon_r$ represent vacuum and relative permittivity, respectively. The time-varying boundary condition on the active contacts is given by the surface integral
\begin{equation}
    \int \limits_{\partial\Omega} \mathbf{J}\cdot \mathbf{n} \diff S = I_0(t),
    \label{eq:boundary_condition}
\end{equation}
where $\mathbf{n}$ is the normal vector to the contact surface.
In the time-dependent model, the current $I_0(t)$ through the active contact varies with time. However, in the static scenario, $I_0(t)$ remains constant, and~(\ref{eq:pde}) simplifies to 
\begin{equation}
    \nabla \cdot (\sigma \nabla u) = 0.
    \label{eq:pde_static}
\end{equation}
Based on the specification of the short Abbott/St. Jude Medical Infinity\texttrademark \ directional lead, the electrode was modeled as a cylinder with a diameter of $\SI{1.27}{mm}$ diameter and contact spacing of $\SI{0.5}{mm}$. 
The lead was enclosed by an encapsulation layer with a thickness of $\SI{0.5}{mm}$. Conductivity in the encapsulation layer was assumed to be homogeneous and set to $\SI{0.18}{S/m}$. Brain tissue in vicinity to the lead was modeled as an $\SI{50}{mm}\times\SI{50}{mm}\times \SI{50}{mm}$ cube with heterogeneous tissue properties. To this end, the $\SI{7}{T}$ ICBM 152 2009a Nonlinear Asymmetric T1 template MRI~\cite{Fonov2009,Fonov2011} was warped to native space and segmented using the inverse transform from previous normalization of native to MNI space. 
Conductivity values for the different tissue types were assigned as follows: gray matter $\SI{0.09}{S/m}$, white matter $\SI{0.06}{S/m}$, and cerebro-spinal fluid $\SI{2.0}{S/m}$. A heterogeneous conductivity model was chosen, because homogeneous models typically underestimate the static VTA. The differences in the electric field norm between homogeneous and heterogeneous tissue conducitivity are illustrated in Fig.~\ref{fig:conductivity}.
The surrounding tissue was modeled as a $\SI{40}{cm}\times\SI{40}{cm}\times \SI{40}{cm}$ volume with a homogeneous conductivity of $\SI{0.1}{S/m}$. 
Voltage- and current-controlled stimulation settings can be investigated by setting boundary conditions to the contact surfaces. Here, the current terminal boundary condition was used on the active contact, while a floating boundary condition was applied to all non-active contacts. 
Fig.~\ref{fig:uni_bipolar} illustrates how unipolar and bipolar stimulation configurations were implemented in COMSOL.

The lead was positioned at the actual DBS lead coordinates obtained from pre- and post-operative CT imaging of an ET patient at Uppsala University Hospital. The imaging data were processed using the open-source software tool Lead-DBS~\cite{Horn2019}. Moreover, the active contacts were chosen based on the clinically used settings (C3-, C4+).

The computation of the static VTA involved applying a threshold of $\SI{150}{V/m}$ to the electric field norm obtained from the solution of the static FEM model.
For the time-dependent computations, the solution obtained from the FEM solver was interpolated at 40 equidistant points along the individual fibers, as indicated by circles in Fig.~\ref{fig:lead_and_tracts}, using a time step of $\SI{5}{\mu s}$. The three fibers were selected based on their proximity to the uppermost ring contact. Specifically, the fiber closest to the ring contact, the one farthest away, and the one positioned in between the two, were chosen.

\subsection{Neuron modeling}\label{sec:neuron}
In order to capture the effect of DBS frequency and pulse-width on fiber activation, the dDRTT and ndDRTT were represented by cable models. 
The effect of extracellular stimulation on a neuron can be described by a multi-compartment Hodgkin-Huxley model, given by the partial differential equation
\begin{equation}
\begin{split}
      I_{m,n} &= g_A \cdot \pdv[2]{u_{m,n}}{x} \\
      &= C_{m,n} \pdv{u_{m,n}}{t} + \sum_{i}^{} g_i (u_m-u_i) + I_{inj}, 
    \label{eq:HH}  
\end{split}
\end{equation}
where $I_{m,n}$ is the total current through the membrane at position $n$.  It is proportional to the axial membrane conductance $g_A$ between successive compartments of the cable. The first term on the right-hand side denotes the capacitive membrane current, which is proportional to the membrane capacitance $C_{m,n}$. The second term describes the contributions of different ion channels with reversal potentials $u_i$. Sodium, potassium, and leak channels were included in the model. The third term $I_{inj}$ captures the influence of both the extracellular electrical potential $u_{e,n}$ and potential neural activity traversing along the fiber.
The model was implemented using the freely available software URDME~\cite{Andersson2018,Drawert2012} for simulating stochastic reaction-diffusion processes. In the context of ion channel gating, URDME employs a stochastic methodology, incorporating kinetic channel models, detailed in~\cite{Senek2017}. The cable length was set to $\SI{8}{mm}$ divided into 40 compartments, whereas the fiber diameter was set to $\SI{2}{\um}$, resulting in compartment lengths approximately 100 times the diameter. The remaining parameters for the neuron model were obtained from~\cite{Bhalla1993}.

\begin{figure}
    \centering
    \includegraphics[scale=0.2]{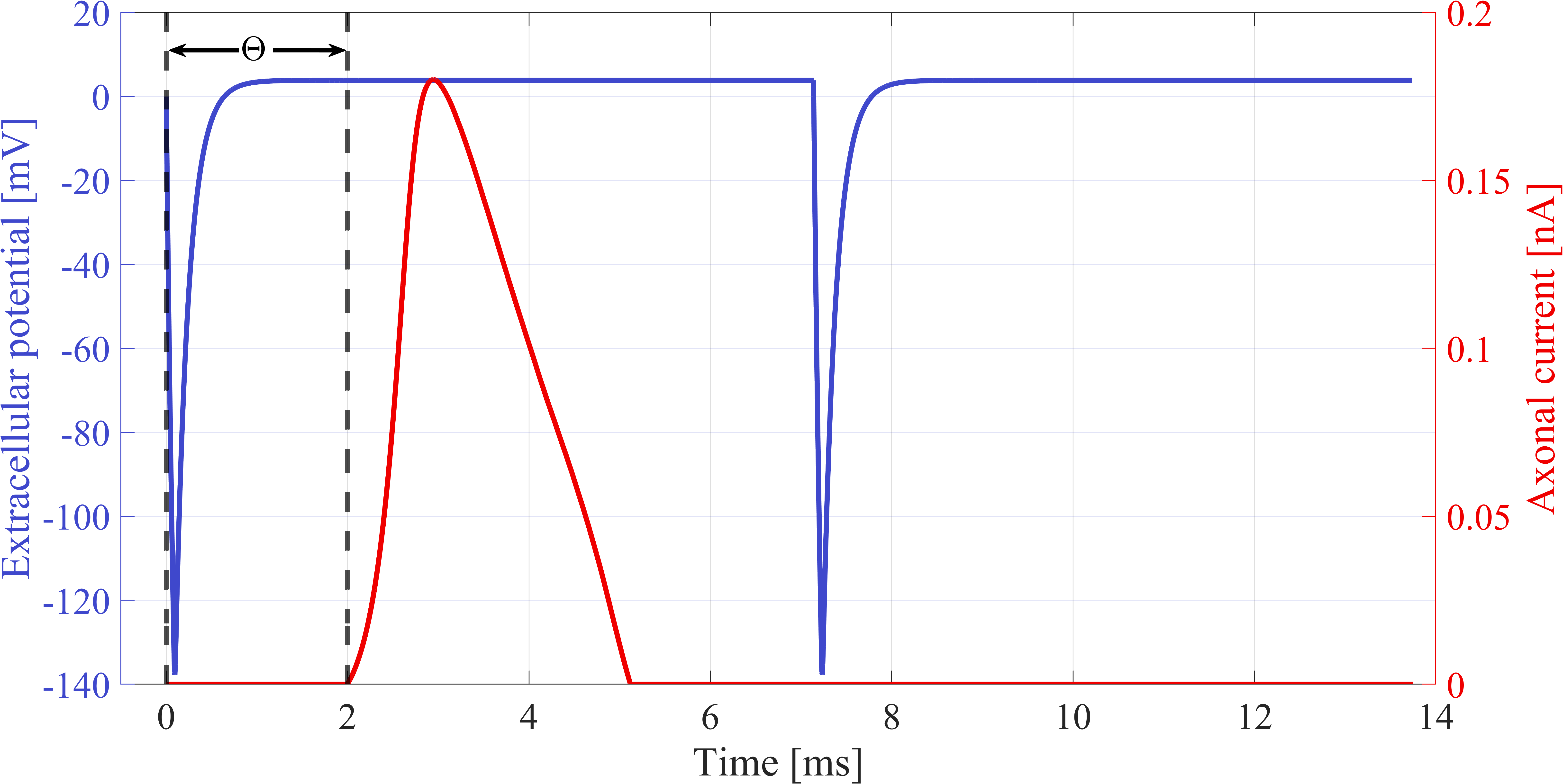}
    \caption{The axonal current and the extracellular potential produced by the DBS lead at one compartment. A phase shift of $\Theta = \SI{2}{ms}$ is showcased.}
    \label{fig:phase_shift}
\end{figure}

A previous study by Andersson \textit{et al.}~\cite{Andersson2018} investigated the role of the phase shift between the input from a neighboring neuron and the DBS pulse on neural firing. 
In the context of this study, the input originating from an adjacent neuron can be more appropriately conceptualized as a neural signal traversing the fiber, which is promoted or suppressed by the DBS pulse. The axonal current as well as the extracellular potential produced by the DBS lead are depicted in Fig.~\ref{fig:phase_shift}.
To explore the influence of the DBS pulse on fiber activation, i.e. whether the neural firing threshold is exceeded and an action potential is generated,  four consequent DBS pulses were examined in conjunction with a solitary axonal input.
The phase shift between the axonal input and the DBS sequence was adjusted in $15$ equally spaced intervals, ranging from zero up to the reciprocal of the stimulus frequency. 
\subsection{Firing scores and binary plots}
\begin{figure}
\begin{subfigure}{\linewidth}
    \centering
    \includegraphics[scale=0.4]{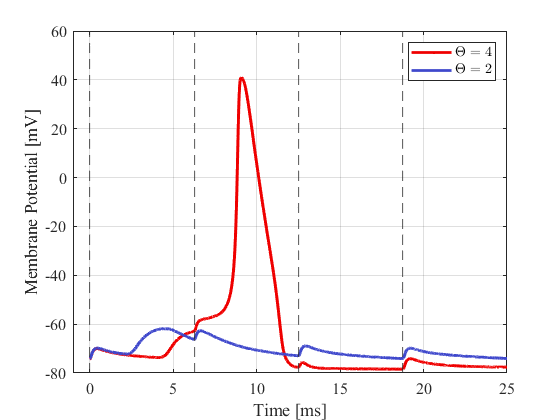}
    \end{subfigure}
    \par\medskip
    \begin{subfigure}{\linewidth}
    \centering
   \includegraphics[width=0.6\textwidth]{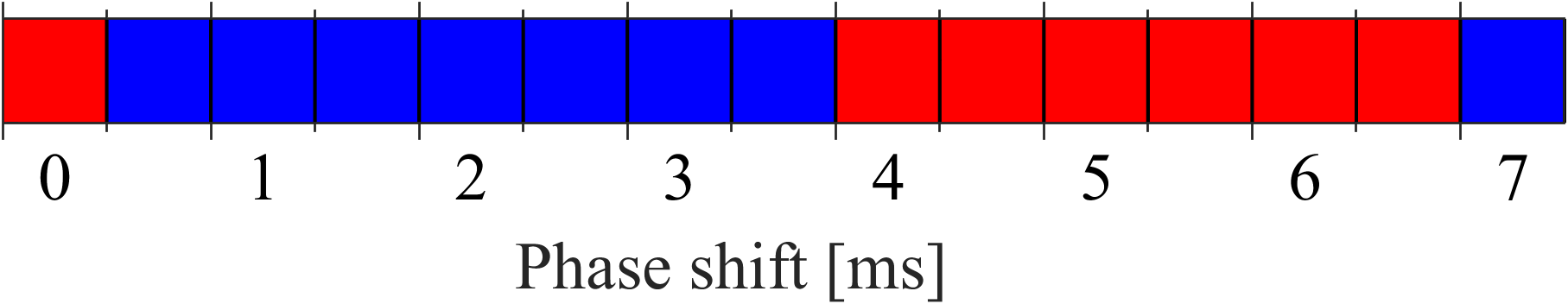}
    \end{subfigure}
        \caption{Top plot: Membrane potential in the first cable compartment for phase shifts $\Theta=2$ (no firing) and $\Theta=4$  (firing) for a unipolar stimulation setting. Bottom plot: Resulting firing pattern for all phase shifts. Vertical dashed lines indicate the initiation of the four DBS pulses.}
    \label{fig:binary_plots}
\end{figure}

Fig.~\ref{fig:binary_plots} displays the neuron model response to a particular DBS configuration, showcasing two distinct phase shifts between the DBS pulses and the axonal current. Additionally, it provides a visual representation of the resulting binary plot across  phase shift values.

The response of the cable model was categorized in a binary manner: absence of firing represented by 0 (blue), and firing represented by 1 (red).
In order to quantify the binary plots in a compact manner, we introduce a firing score given by the ratio of the number of phase shifts for which firing is achieved to the total number of phase shifts examined, i.e.

\begin{equation}
    \text{Firing Score}= \frac{\text{Number of Phase Shifts with Firing}}{\text{Total Number of Phase Shifts}}  .
    \label{eq:firing_score}
\end{equation}
The total number of phase shifts was set to 15, whereas the step size of phase shifts varied with stimulation frequency.

\section{Results}\label{sec:results}
\subsection{Role of static modeling}
\begin{figure}
    \centering
    \begin{subfigure}{0.47\linewidth}
      \includegraphics[scale=0.3]{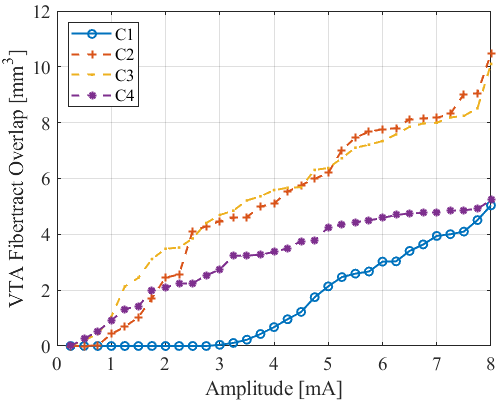}  
      \subcaption[]{dDRTT}
    \end{subfigure}
    \begin{subfigure}{0.47\linewidth}
    \includegraphics[scale=0.3]{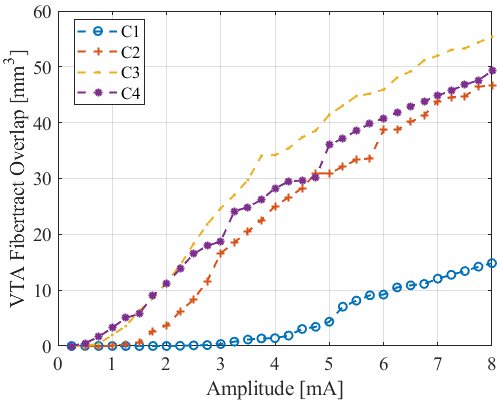}  
    \subcaption[]{ndDRTT}
    \end{subfigure}
    
    \caption{Overlap of the static VTA with the dDRTT and ndDRTT with increasing amplitude. The upper ring contact (C4-) was active and the VTA threshold was set to $\SI{150}{V/m}$.}
    \label{fig:amplitudes_VTA}
\end{figure}

Initially, the static model, i.e. \eqref{eq:pde} with constant boundary conditions, was solved for unipolar configurations of all contact rows (C1 to C4). Fig.~\ref{fig:amplitudes_VTA} illustrates the relationship between amplitude and the overlap between the static VTA and the fiber tracts. For the patient considered in this study, there is indeed an overlap between the conventionally computed VTA and both sections of the fiber tracts. However, the overlap with the dDRTT is significantly smaller in comparison to that with the ndDRTT. Notably, the largest overlap can be observed for the contact rows that correspond to the clinically active contacts (C3, C4). This observation suggests that the traditional VTA models may still prove useful in identifying the most effective contact combination for covering the fiber tract.

\subsection{Effect of dynamical stimulation parameters}
\begin{figure}
\begin{subfigure}{\linewidth}
    \centering
    \includegraphics[width=0.325\linewidth]{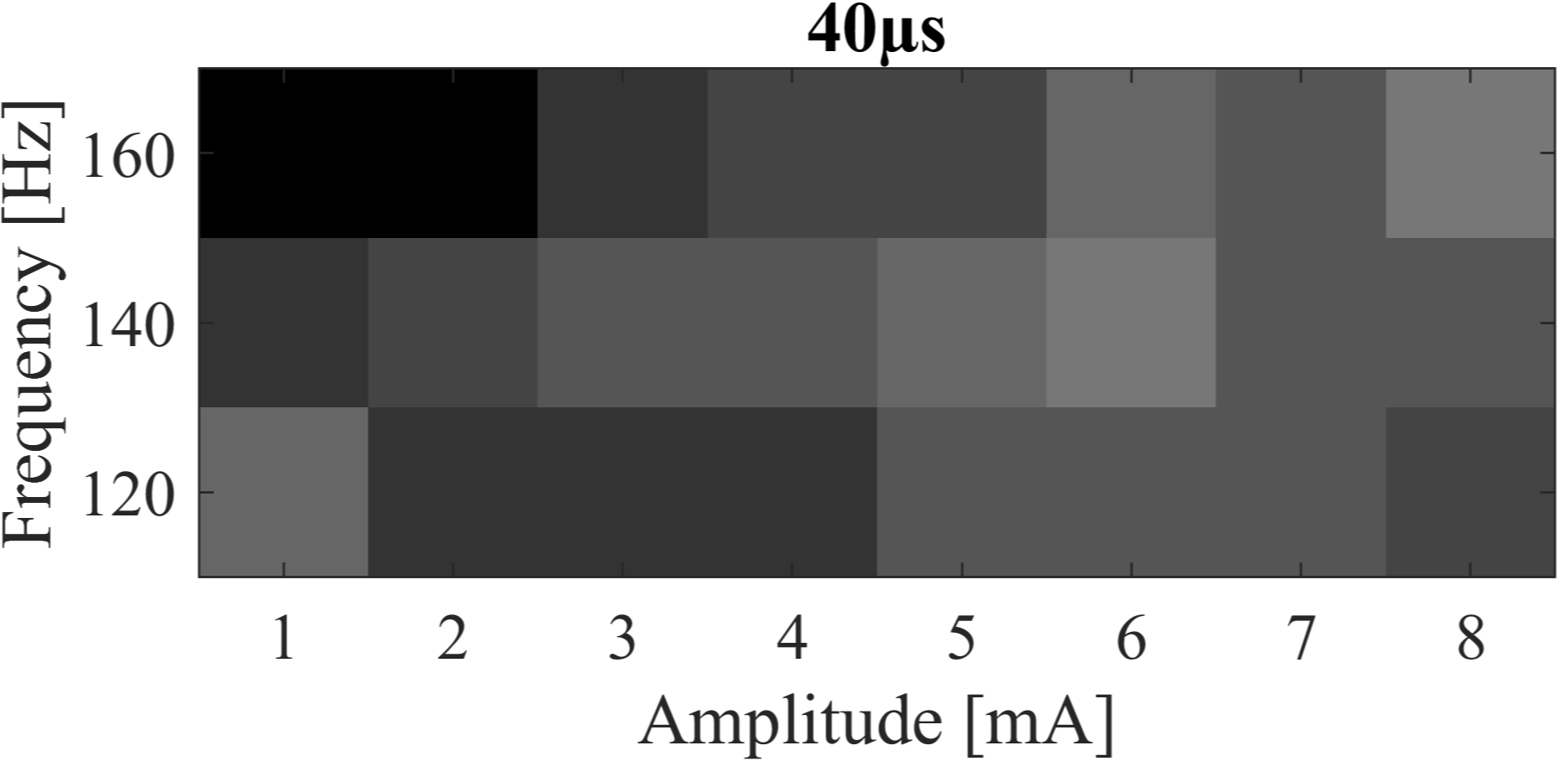}
    \includegraphics[width=0.29\linewidth]{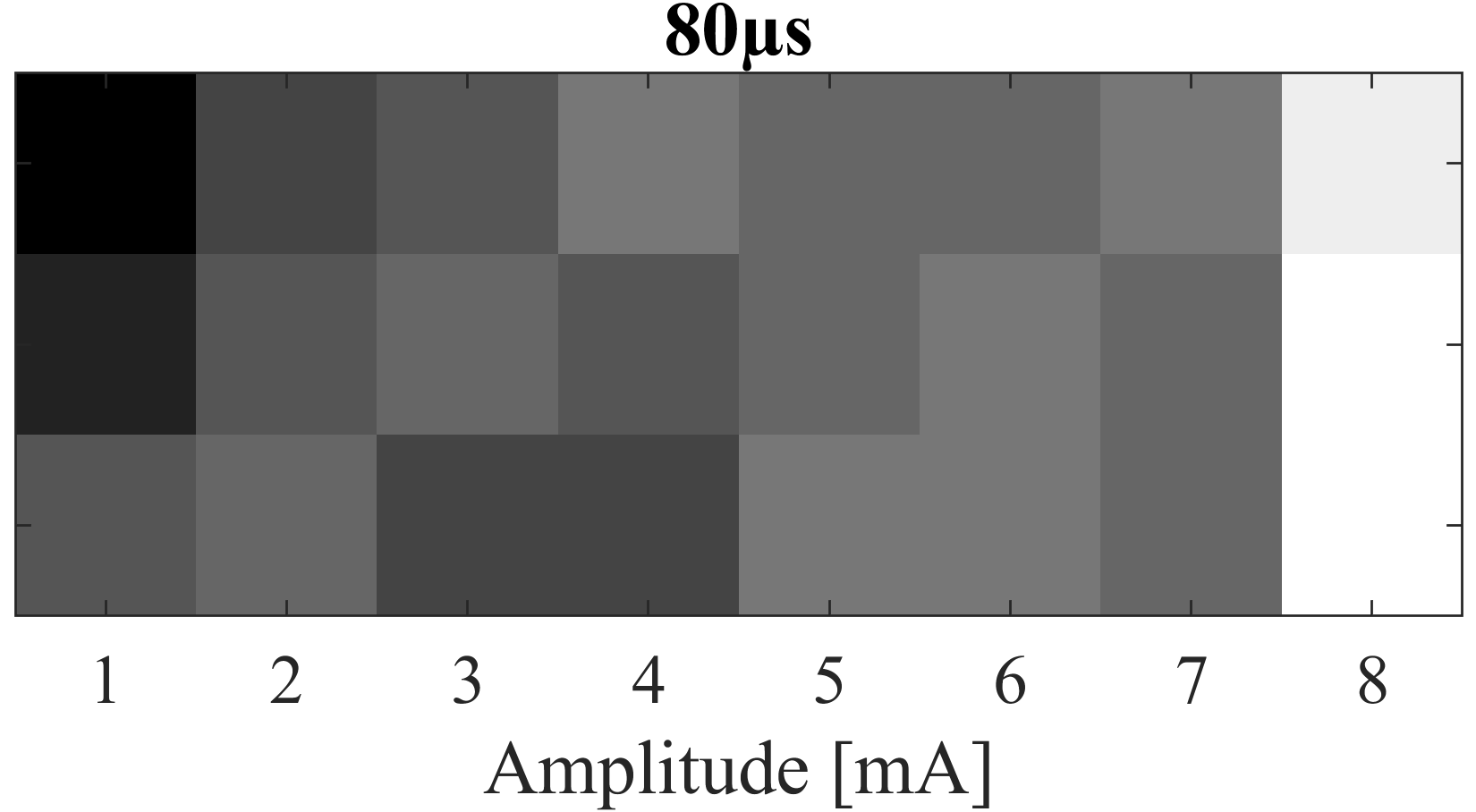}
    \includegraphics[width=0.35\linewidth]{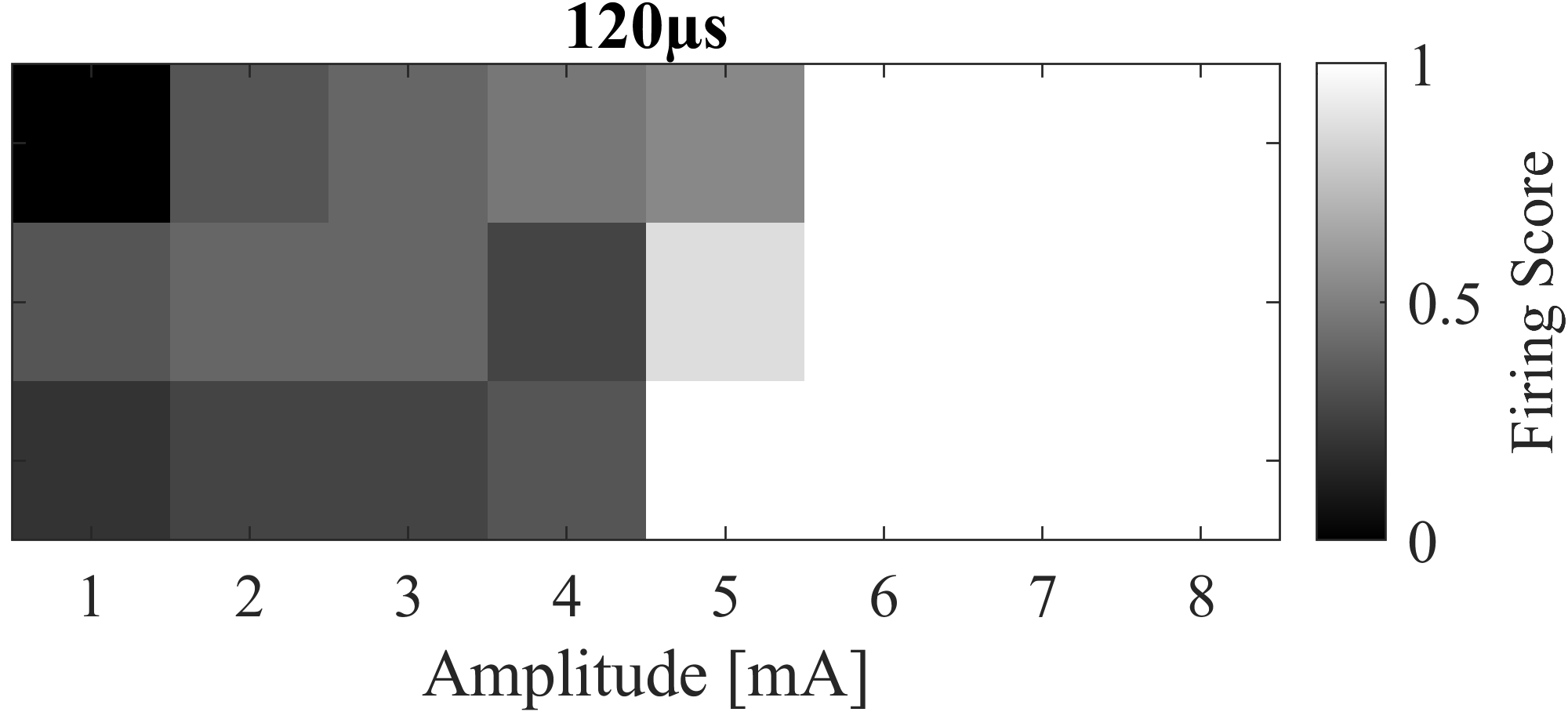}
\end{subfigure}
\par\medskip
\begin{subfigure}{\linewidth}
        \centering
    \includegraphics[width=0.325\linewidth]{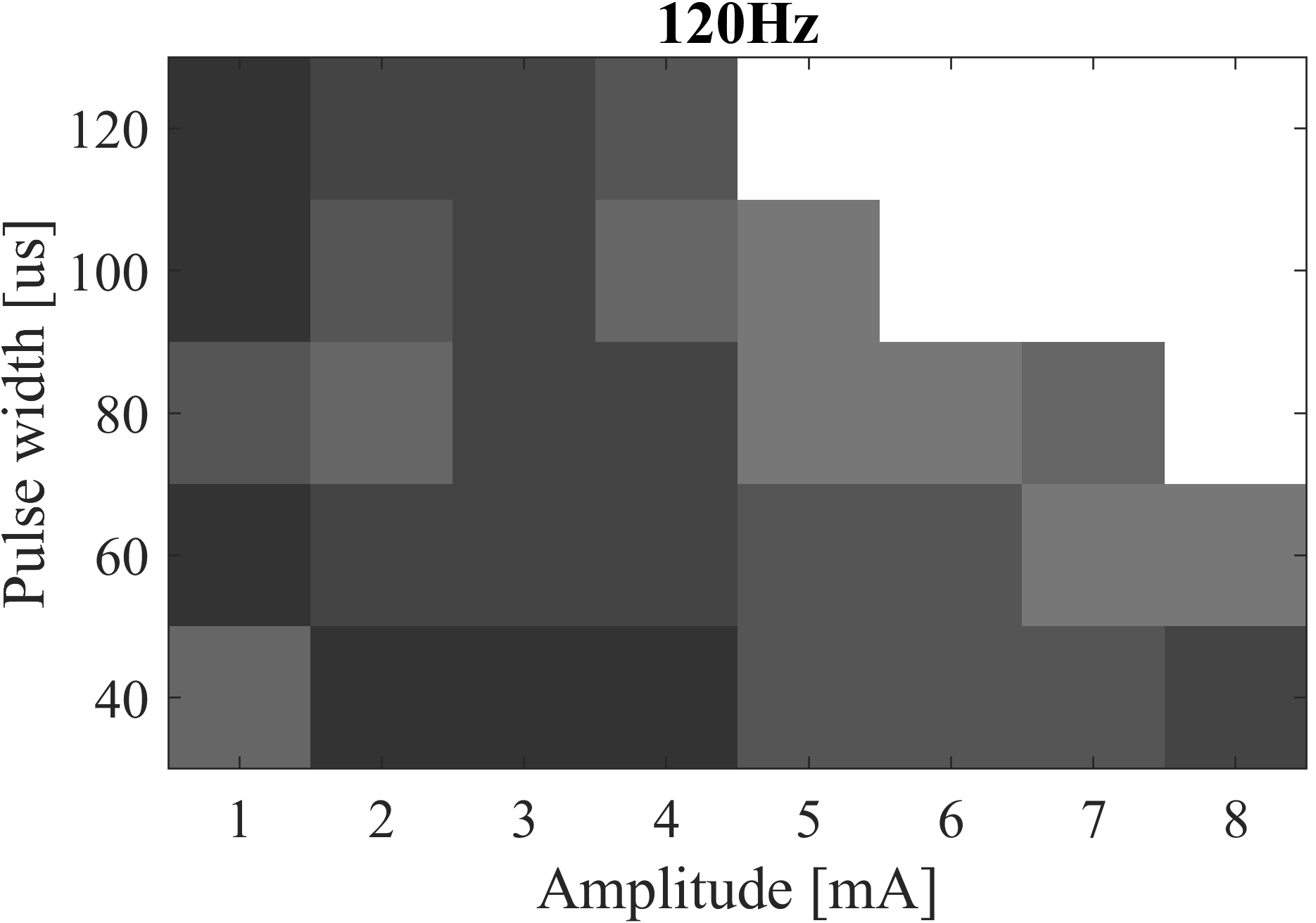}
    \includegraphics[width=0.29\linewidth]{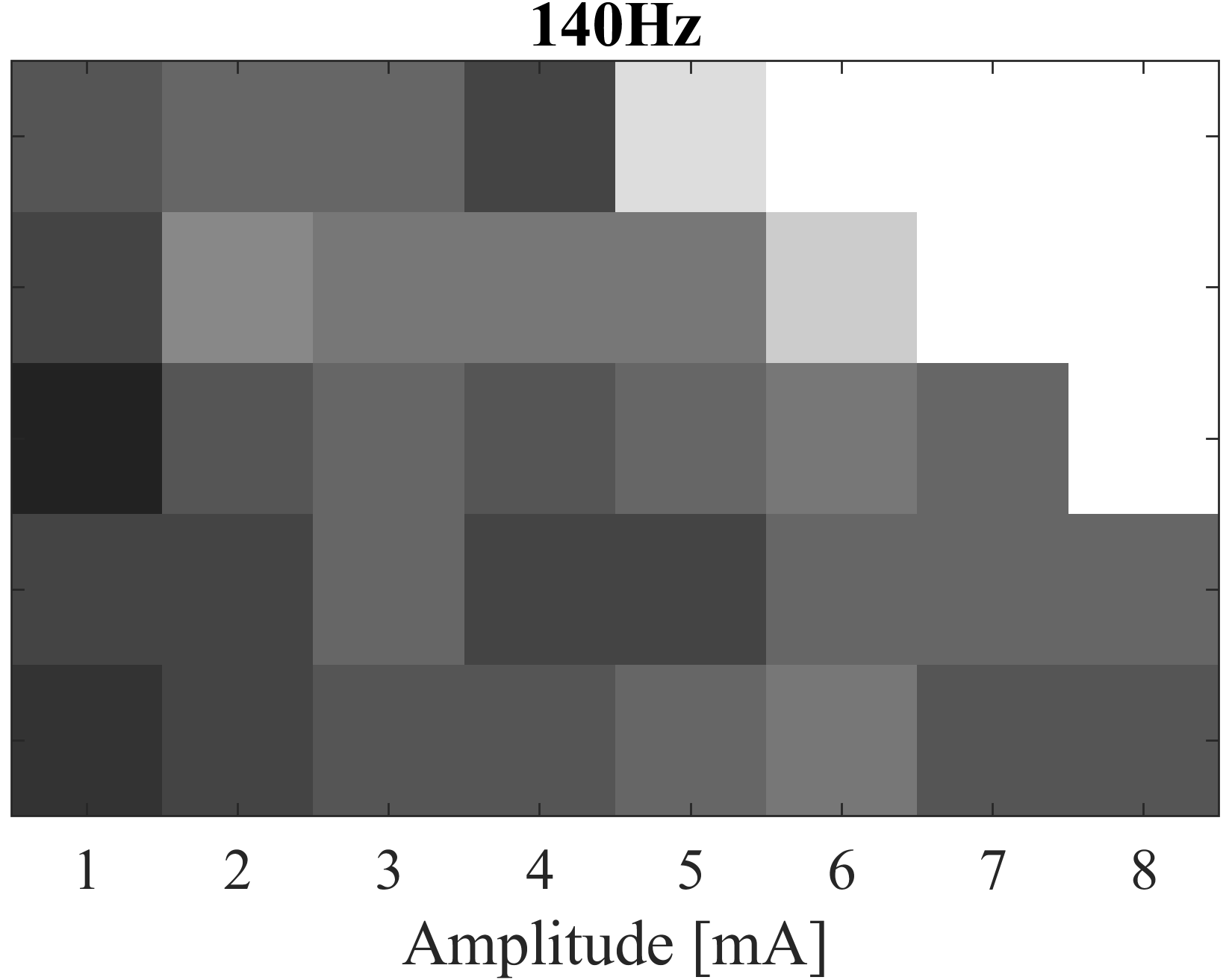}
    \includegraphics[width=0.35\linewidth]{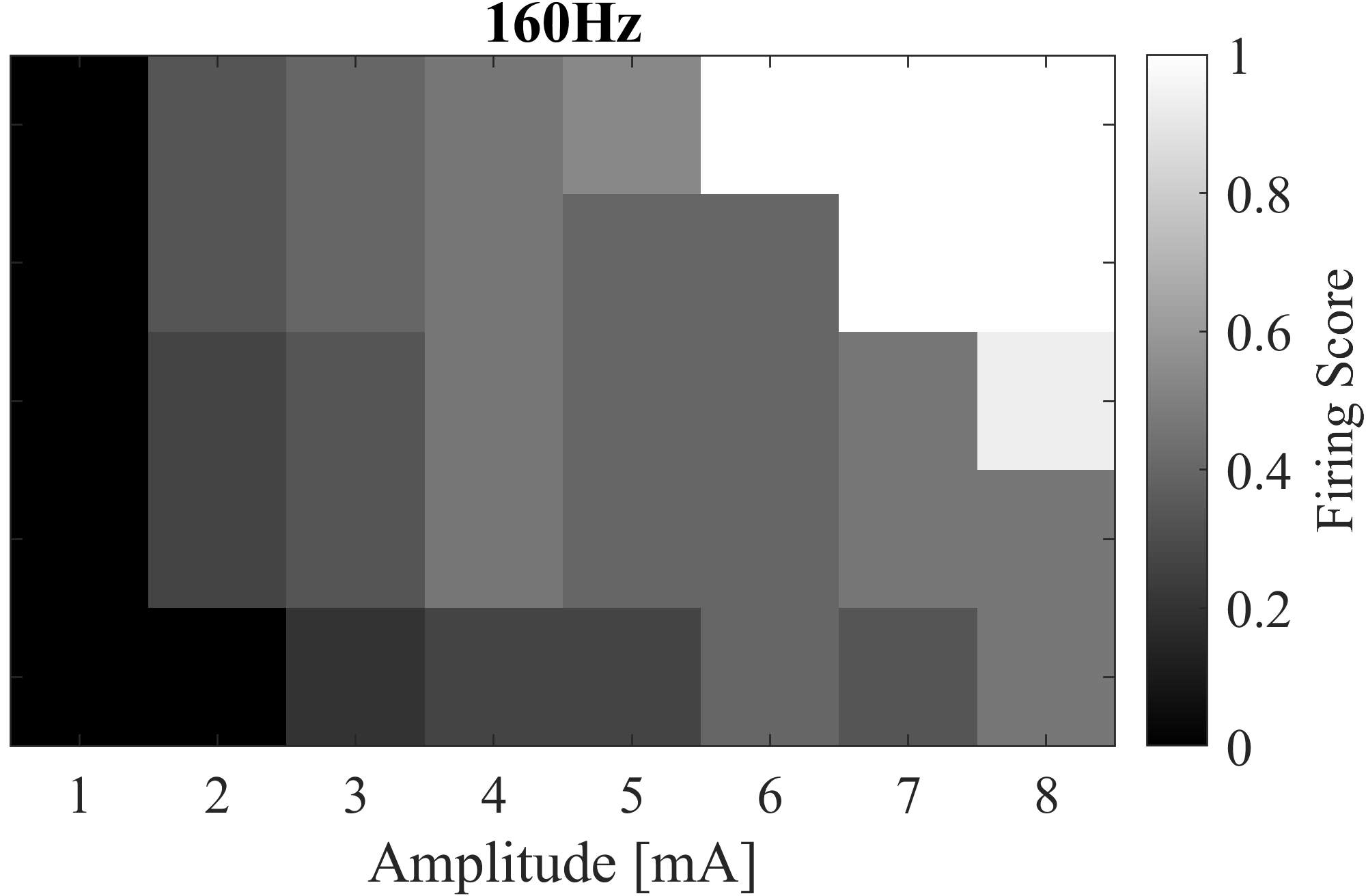}
\end{subfigure}
    \caption{Firing scores for different  pulse widths (top) and frequencies (bottom) versus amplitudes under unipolar (C4-) stimulation settings. The grey-scale represents firing score \eqref{eq:firing_score}, ranging from 0 (no firing) to 1 (firing occurring at all phase shifts). The cable model was positioned on the dDRTT fiber with shortest distance to the active contact.}
    \label{fig:unipolar_gridded}
\end{figure}

The effect of pulse width and frequency in unipolar stimulation settings were investigated for the fiber closest to the lead, marked in Fig.~\ref{fig:lead_and_tracts}.
Fig.~\ref{fig:unipolar_gridded} displays how changes in amplitude and pulse widths impact the firing score under unipolar settings.
Clearly, a specific amplitude is necessary to induce any modulation effect. Longer pulses (higher values of pulse widths) are more effective in facilitating firing, whereas the influence of frequency is less distinct. 
It is important to note that certain variations arise due to the statistical nature of the used neuron model, see Section~\ref{sec:neuron}.

\subsection{Bipolar stimulation efficacy}
\begin{figure}
\begin{subfigure}{\linewidth}
    \centering
    \includegraphics[width=0.325\linewidth]{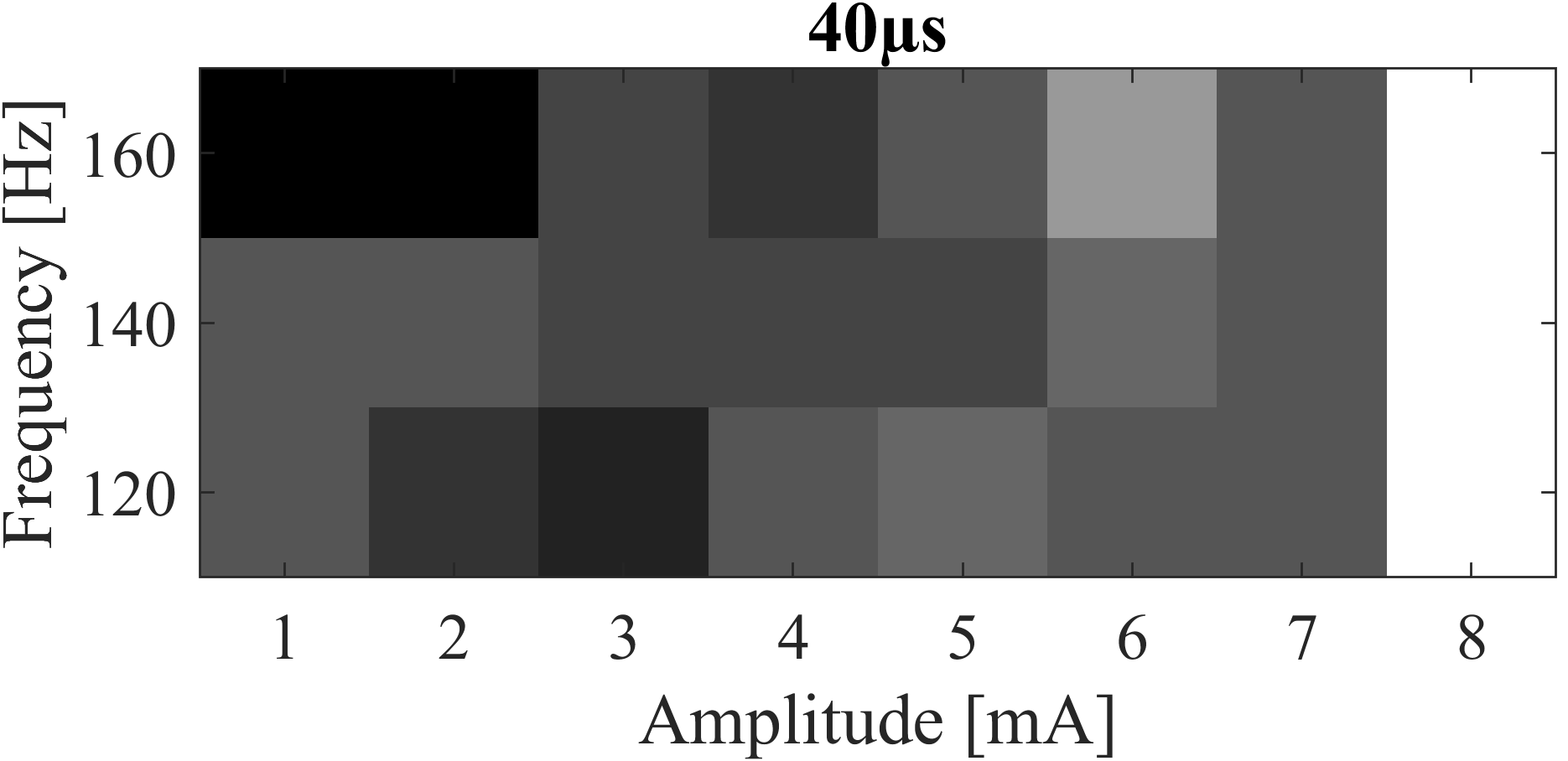}
    \includegraphics[width=0.29\linewidth]{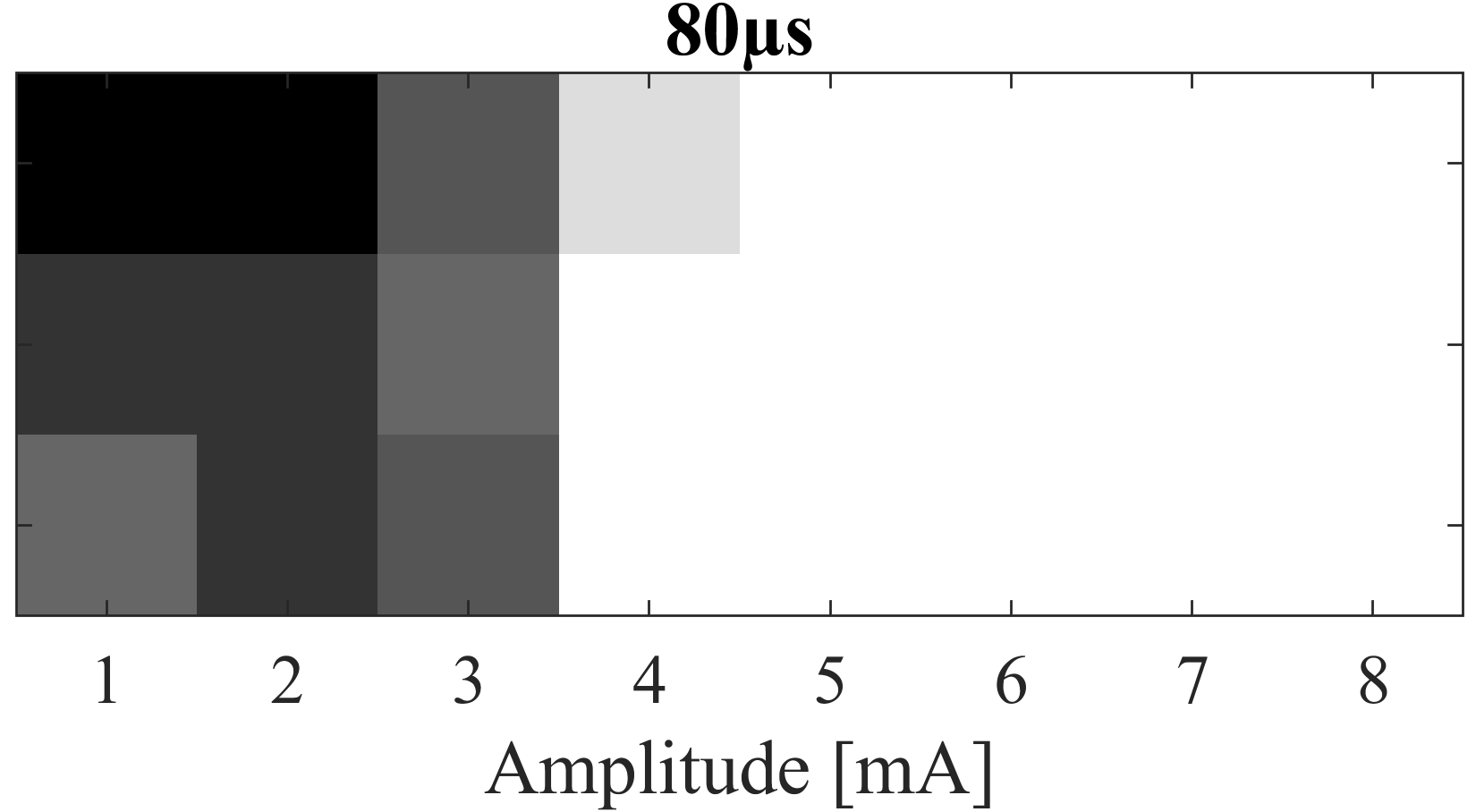}
    \includegraphics[width=0.35\linewidth]{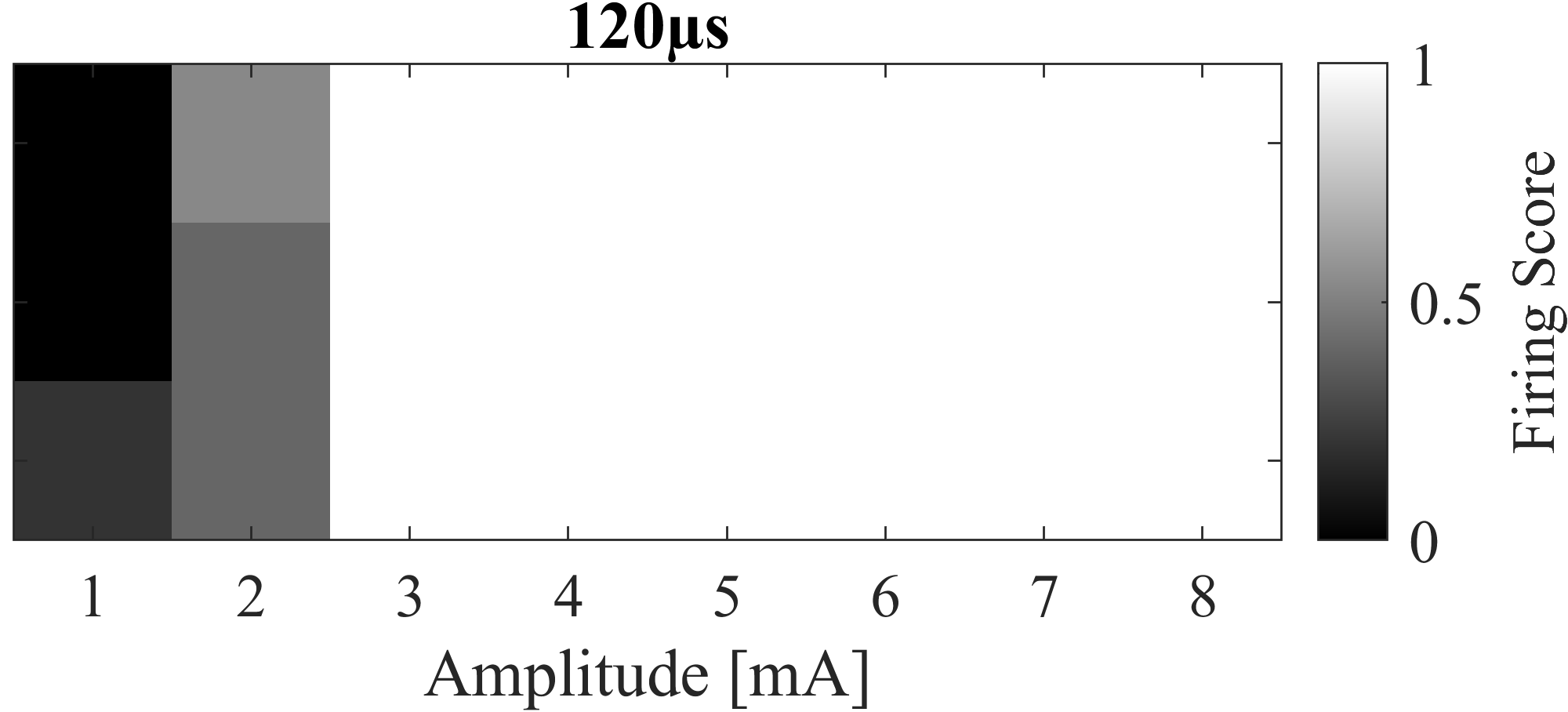}
\end{subfigure}
\par\medskip
\begin{subfigure}{\linewidth}
        \centering
    \includegraphics[width=0.325\linewidth]{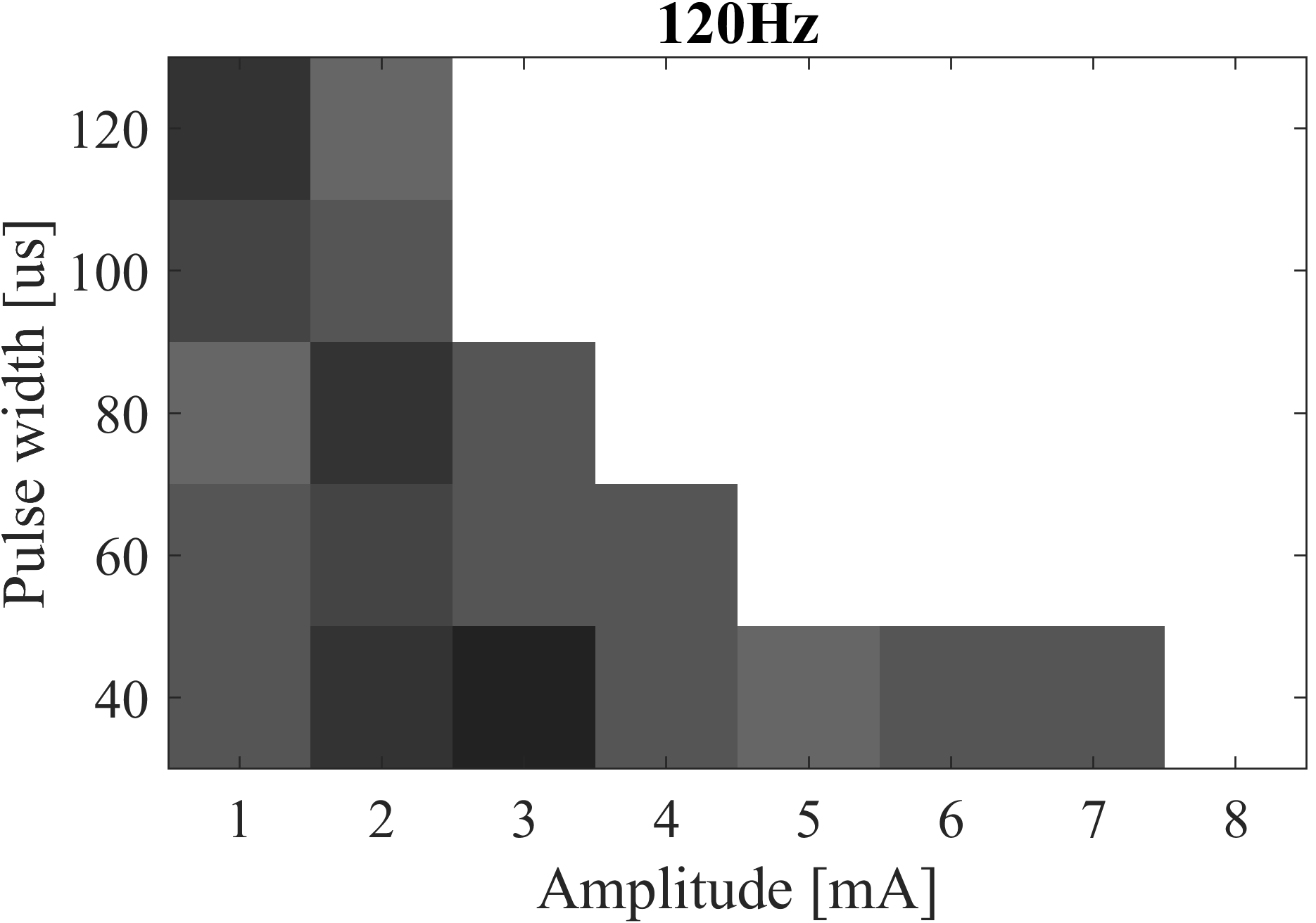}
    \includegraphics[width=0.29\linewidth]{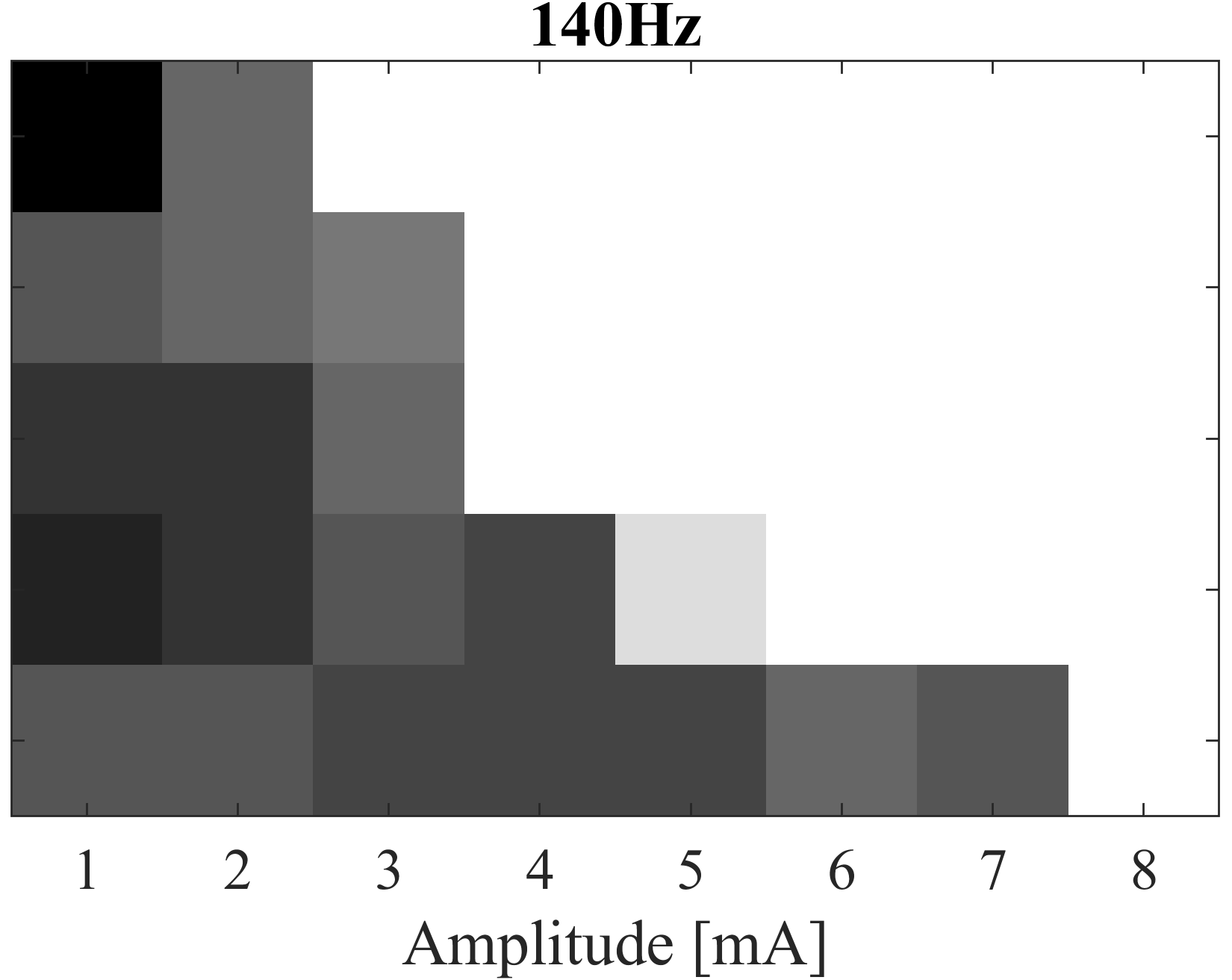}
    \includegraphics[width=0.35\linewidth]{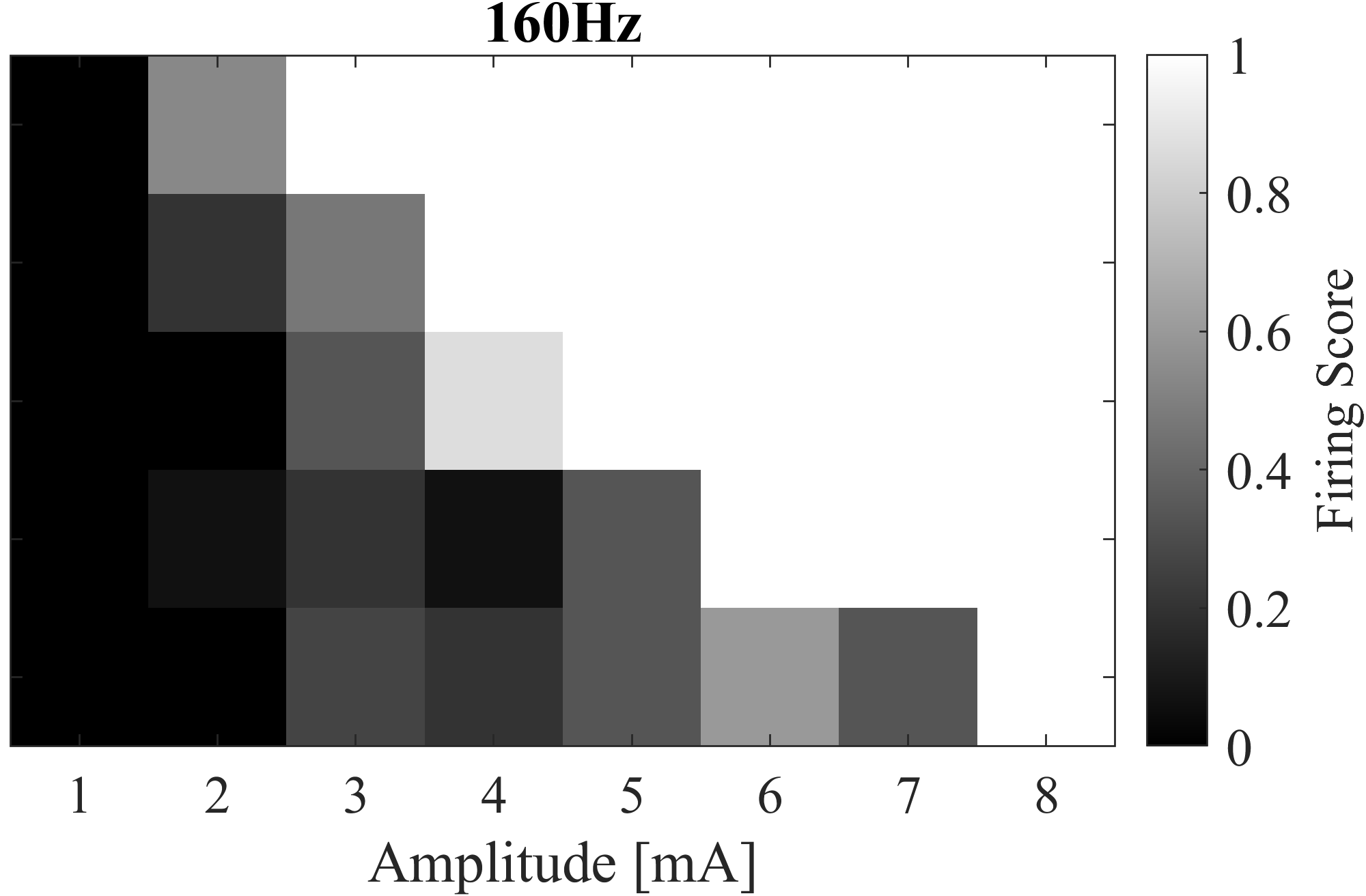}
\end{subfigure}
    \caption{Firing scores for different pulse widths (top) and
frequencies (bottom) versus amplitudes under bipolar (C4-,~C3+) stimulation settings. The grey-scale represents firing
score (5), ranging from 0 (no firing) to 1 (firing occurring
at all phase shifts). The cable model was positioned on the
dDRTT fiber with shortest distance to the active contact.}
    \label{fig:bipolar_gridded}
\end{figure}

The results for bipolar stimulation, as depicted in Fig.~\ref{fig:bipolar_gridded}, yielded analogous findings to those presented in Fig.~\ref{fig:unipolar_gridded} for unipolar stimulation. 
 While increased pulse widths resulted in increased firing rates in both cases, it is noteworthy that, in the context of bipolar stimulation, lower frequencies produce a more pronounced modulation effect on neuron firing.
In general, the selected bipolar (C4-, C3+) configuration induced a greater degree of firing compared to the unipolar (C4-) configuration.

\begin{figure}[thpb]
      \centering
      \includegraphics[width=0.35\textwidth]{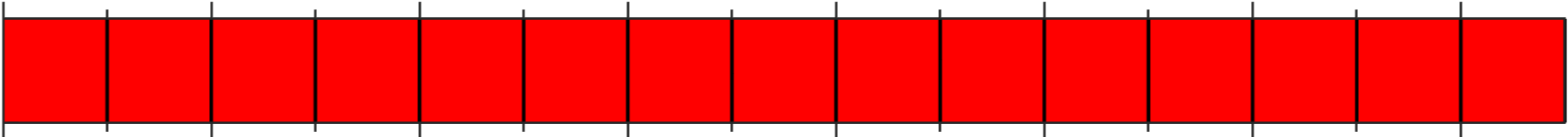}
      \includegraphics[width=0.35\textwidth]{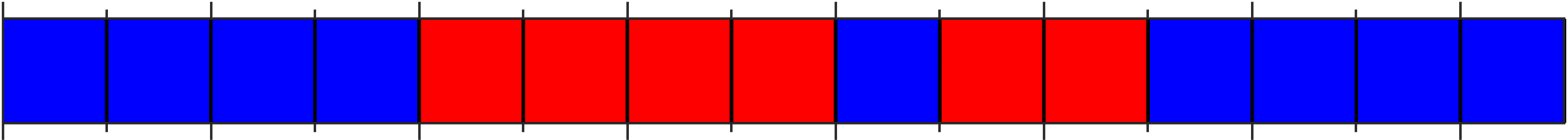 }
      \includegraphics[width=0.35\textwidth]{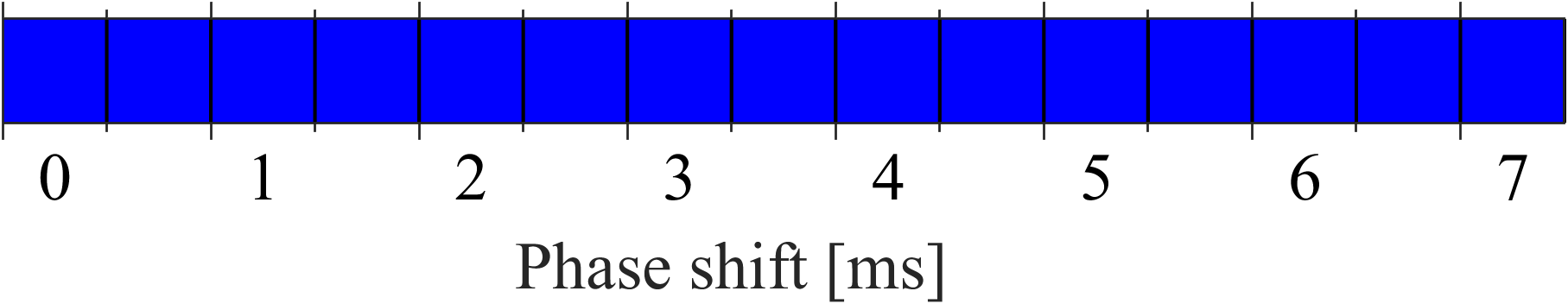}
      \caption{Binary firing plots for three individual fibers of the dDRTT under a single bipolar setting (C3-, C4+) with a frequency of \SI{140}{Hz}, a pulse width of \SI{90}{us}, and an amplitude of \SI{3}{mA}. The corresponding fibers are highlighted in red in Fig.~\ref{fig:lead_and_tracts}. The plots from top to bottom depict an increasing distance between the fiber and the DBS lead.}
      \label{fig:bipolar_3fibs}
   \end{figure}

Fig.~\ref{fig:bipolar_3fibs} illustrates the firing patterns of the three distinct fibers within the dDRTT (see Fig.~\ref{fig:lead_and_tracts}) under a single bipolar setting.
As the distance between the lead and the fiber increases, the influence of the DBS stimulus on fiber activation diminishes, as one would intuitively expect.

\begin{figure}
\captionsetup[subfigure]{labelformat=empty}
  \centering
  \begin{subfigure}{0.51\linewidth}
    \centering
    \begin{minipage}{\textwidth}
      \centering
      \includegraphics[width=\linewidth]{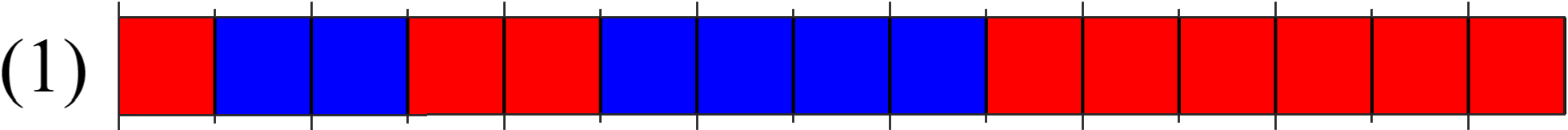}
    \end{minipage}
    \hfill
    \begin{minipage}{\textwidth}
     \centering
      \includegraphics[width=\linewidth]{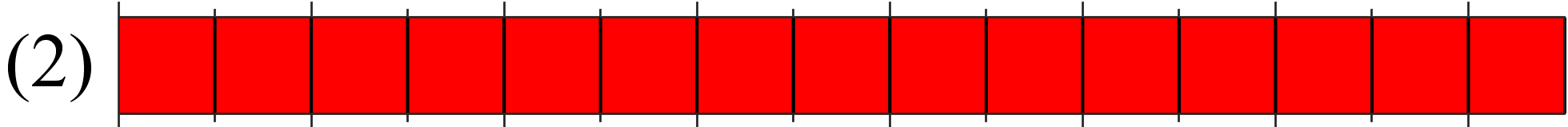}
    \end{minipage}
    \hfill
    \begin{minipage}{\textwidth}
    \centering 
      \includegraphics[width=\linewidth]{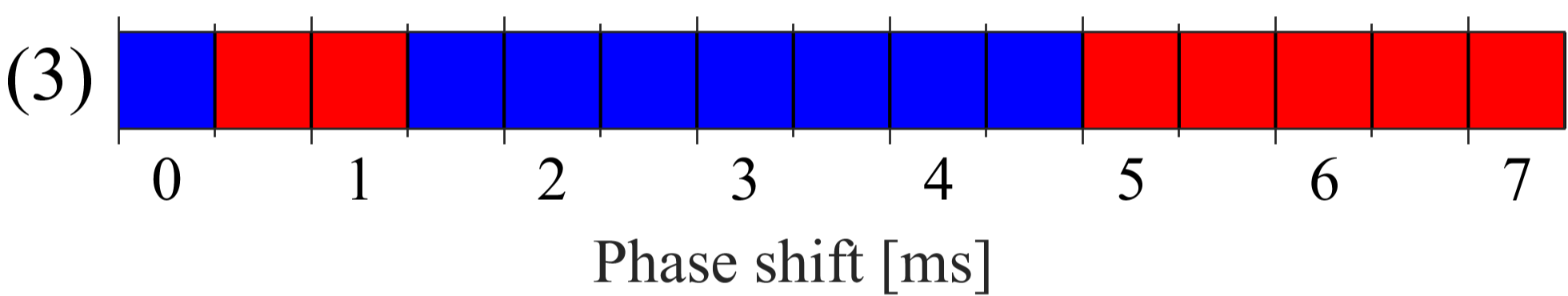}
    \end{minipage}
    \caption{(a) dDRTT}
  \end{subfigure}
  \hfill
  \begin{subfigure}{0.47\linewidth}
    \centering
    \begin{minipage}{\textwidth}
      \centering
      \includegraphics[width=\linewidth]{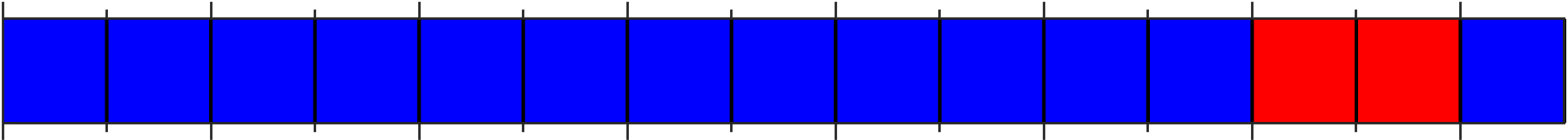}
    \end{minipage}
    \hfill
    \begin{minipage}{\textwidth}
      \centering
      \includegraphics[width=\linewidth]{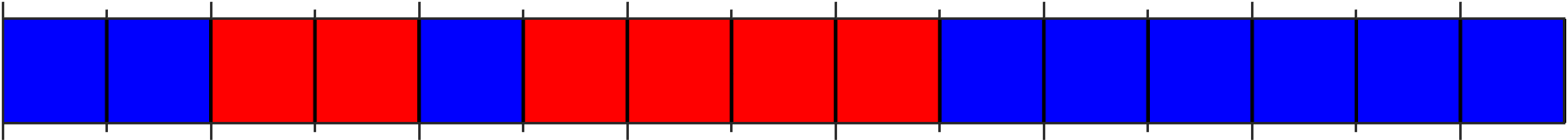}
    \end{minipage}
    \hfill
    \begin{minipage}{\textwidth}
      \centering
      \includegraphics[width=\linewidth]{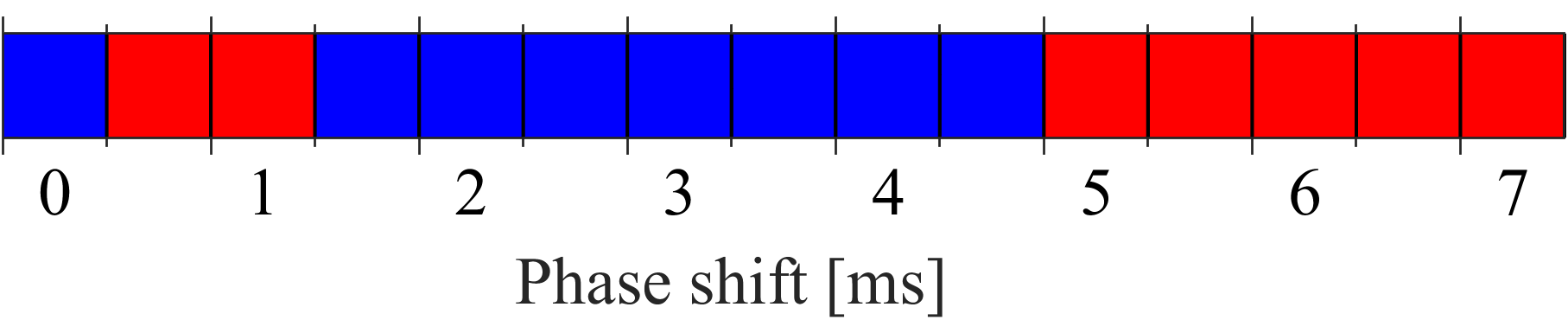}
    \end{minipage}
    \caption{(b) ndDRTT}
  \end{subfigure}

    \begin{subfigure}{0.51\linewidth}
    \centering
    \begin{minipage}{\textwidth}
      \centering
      \includegraphics[width=\linewidth]{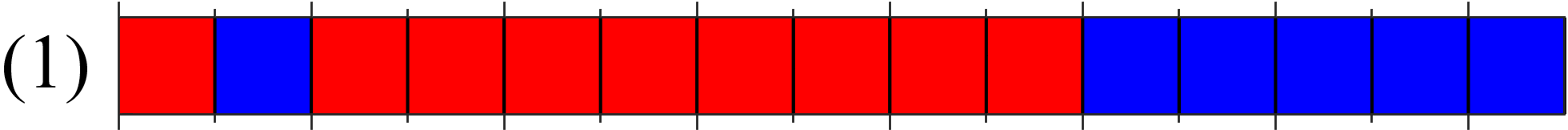}
    \end{minipage}
    \hfill
    \begin{minipage}{\textwidth}
     \centering
      \includegraphics[width=\linewidth]{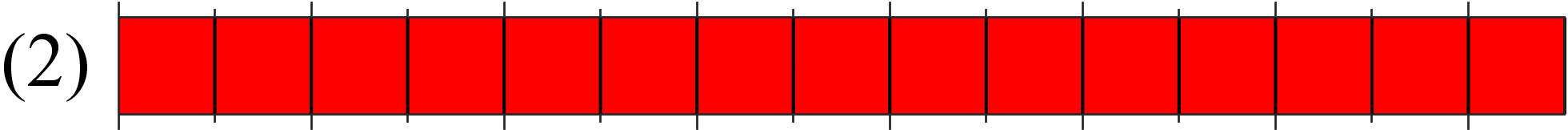}
    \end{minipage}
    \hfill
    \begin{minipage}{\textwidth}
    \centering 
      \includegraphics[width=\linewidth]{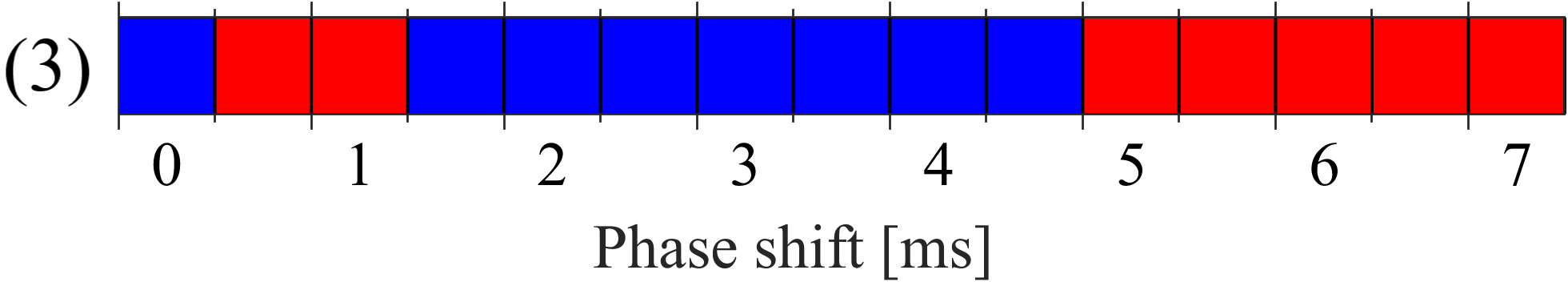}
    \end{minipage}
    \caption{(c) dDRTT flipped}
  \end{subfigure}
  \hfill
  \begin{subfigure}{0.47\linewidth}
    \centering
    \begin{minipage}{\textwidth}
      \centering
      \includegraphics[width=\linewidth]{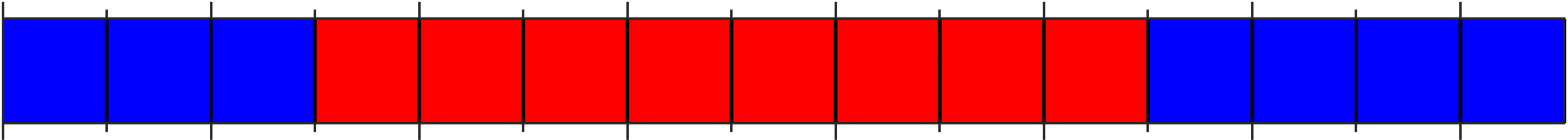}
    \end{minipage}
    \hfill
    \begin{minipage}{\textwidth}
      \centering
      \includegraphics[width=\linewidth]{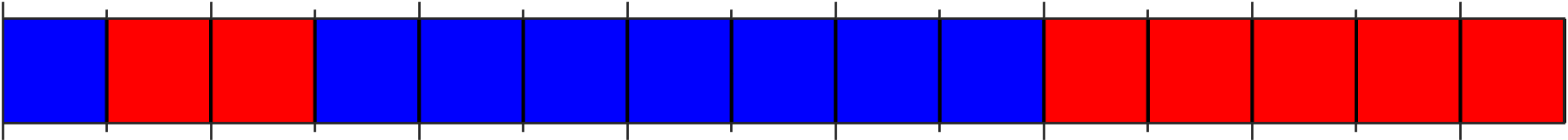}
    \end{minipage}
    \hfill
    \begin{minipage}{\textwidth}
      \centering
      \includegraphics[width=\linewidth]{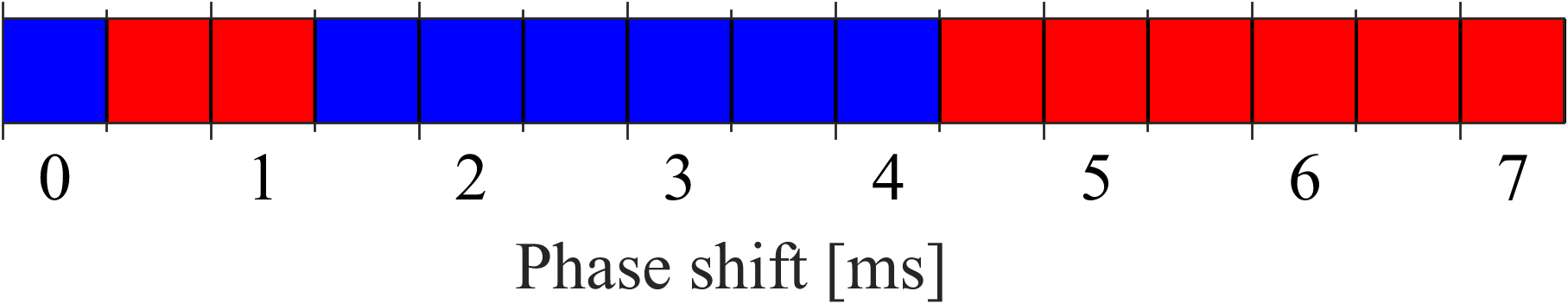}
    \end{minipage}
    \caption{(d) ndDRTT flipped}
  \end{subfigure}
  \caption{ 
  Binary firing patterns for bipolar settings with reversed polarity, unipolar setting, in both the dDRTT and the ndDRTT. Subplots (a),(b): original neural traffic direction, (c),(d): flipped neural traffic neural direction. Only the closest to the lead fibers are considered.  Top plots (1):  configuration of (C3-, C4+) polarity; Center plots (2): reversed (C4-, C3+) polarity; Bottom plots (3): unipolar configuration (C3-, C4-). The dynamic stimulation parameters for all plots: frequency \SI{140}{Hz}, pulse width \SI{90}{us},  amplitude \SI{3}{mA}.}
  \label{fig:bipolar_at_tracts}
\end{figure}

In Fig.~\ref{fig:bipolar_at_tracts}, the binary firing plots for two bipolar settings of reversed polarity and a corresponding unipolar setting are provided for both the dDRTT and the ndDRTT. While one of the bipolar settings demonstrates a firing pattern resembling that of the unipolar setting, reversing the polarity significantly increases the likelihood of modulating neural traffic to induce firing. Furthermore, the bipolar settings induce notably higher firing activity in the dDRTT compared to the ndDRTT, demonstrating the ability to selectively influence firing patterns in these two tracts.
Additionally, subplots in Fig.~\ref{fig:bipolar_at_tracts} (c) and (d) depict a similar scenario but with a reversed cable model, where the direction of neural traffic is flipped. The results show that bipolar configurations can trigger firing for nearly opposite phase shifts when neural traffic direction is reversed.

\section{Discussion}

In this paper, the influence of dynamic parameters and polarity of DBS configurations on fiber activation were examined in a realistic computational model.
\paragraph*{Role of static modeling}
Static VTA computations can prove valuable in finding the most efficient contact configuration and in determining an amplitude threshold for stimulation effects, a prerequisite for any subsequent adjustments of frequency, pulse width, and polarity. However, static models are of limited utility in ET, where dynamical DBS parameters have significant influence on symptom management. Hence, the need for more sophisticated models arises to capture and explicate the effect of dynamical DBS parameters. Being solved in space and time, these models are more computationally intensive and require high-performance hardware to be practically implemented. Running the high resolution spatio-temporal FEM model once requires about one hour on a standard computer, whereas a single run of the neuron model is completed in approximately two minutes.

The modeling framework introduced in this study does not rely on the theories regarding the underlying mechanisms of DBS and is equally useful in investigating both inhibition and excitation of neural signals, as well as disruption of abnormal information flow. In this study, however, the focus was put on how DBS enhances neural signals to induce firing.
\paragraph*{Dynamic stimulation parameters}
The presented simulation results indicate that longer pulse widths enable increased alignment between a traversing axonal current and the DBS pulse, leading to a more pronounced modulation effect. These observations are consistent with prior research \cite{Anderson2019}, yet embedding the concept in a more realistic setting.
In clinical practice, it is observed that shorter pulse widths require higher amplitudes to achieve similar symptom relief~\cite{Koeglsperger2019}, a pattern that aligns well with the simulation results of this study, see Fig.~\ref{fig:unipolar_gridded}. Conversely, longer pulse widths are typically linked to an increased likelihood of experiencing side effects~\cite{Koeglsperger2019}.
Considering the modeling configuration, one might expect that higher stimulation frequencies would allow for a shorter engagement intervals with axonal currents, potentially resulting in a more pronounced modulation effect, whether through suppression or enhancement.  
Nonetheless, the results in Section~\ref{sec:results} suggest that in the case of bipolar stimulation, lower DBS frequencies may have a higher chance of promoting neural firing. It is important to note, however, that all the frequencies examined in this study are still categorized as high-frequency DBS~\cite{Vallabhajosula2015}.

\paragraph*{Unipolar vs bipolar}
The presented findings highlight that bipolar settings can yield distinct effects on axonal currents, contingent upon the polarity and direction of neural traffic. This emphasizes the significance of exploring bipolar settings within clinical practice and stresses the need to devise methods for modeling their effects to enhance parameter tuning efficiency.
In future work, we have to objectives. First, we intend to relate the firing patterns presented in this paper with symptom severity in patients. Second, we aim to define a VTA for bipolar stimulation configurations based on the conceptual framework presented in this paper. 

The study is subject to several limitations. These include uncertainties in parameters such as cable diameter, axonal input characteristics, and the assumption of a single axonal input. Additionally, the  actual conductivity of the tissue  around the lead and encapsulation layer thickness may affect the direct translation of amplitudes used in the simulations to real-world settings.

\section{Conclusions}
The paper presents a computational modeling framework that simulates the impact of dynamic stimulation parameters and bipolar settings on neural modulation in DBS.
The electrode position, brain tissue conductivity, as well as the stimulation target are patient-specific, whereas the neural dynamics are described by an established advanced mathematical model.
This research paves the way for future work correlating these findings with patient symptom profiles, offering the potential for a more precise definition of VTAs for bipolar stimulation configurations in clinical practice.


\section*{ACKNOWLEDGMENT}
The computations were enabled by resources provided by the National Academic Infrastructure for Supercomputing in Sweden (NAISS) and the Swedish National Infrastructure for Computing (SNIC) at UPPMAX partially funded by the Swedish Research Council through grant agreements no. 2022-06725 and no. 2018-05973.

Moreover, the authors would like to thank Stefan Engblom, Department of Information Technology, Uppsala University, Sweden, for his help with the mathematical model of the neuron and the software URDME used in this paper.

\bibliographystyle{IEEEtran}
\bibliography{IEEEabrv,references}
\addtolength{\textheight}{-12cm}

\end{document}